\documentclass[
aps,
prx,
superscriptaddress,
showpacs,
showkeys,
twocolumn,
numbers,
longbibliography,
nofootinbib,
10pt]
{revtex4-2}

\usepackage[T1]{fontenc} % alphabets to prepare
\usepackage{bigints}
\usepackage[utf8]{inputenc} % input encoding
\usepackage[english]{babel} % language(s)
\usepackage{graphicx, xcolor}
\usepackage{amsfonts, amssymb, amsmath, mathrsfs, calc, array, mathtools}
\usepackage[separate-uncertainty=true]{siunitx}
\usepackage{hyperref}
\usepackage{abraces}
\usepackage{bbm}
\usepackage{tikz}
\usetikzlibrary{quantikz2}
\usepackage{listings}
\usepackage{tikz-cd}
\usepackage{xcolor}

\def\ket#1{\mathinner{|{#1}\rangle}}

\def\ontop#1#2{\setbox0\hbox{#2}\copy0\llap{\raise\ht0\hbox{#1}}}

\definecolor{darkblue}{rgb}{0,0,0.93} 
\definecolor{darkred}{rgb}{0.8,0,0} 
\definecolor{darkgreen}{rgb}{0,0.7,0} 
\hypersetup{
colorlinks=true, 
linkcolor=darkblue, 
citecolor=darkred, 
filecolor=darkblue, 
urlcolor=darkblue
}

\begin{document}
%\title{On the numerical limitations of dual Koopman–von Neumann embeddings for quantum computation of conservative nonlinear ordinary differential equations }
\title{On the numerical limitations of dual Koopman–von Neumann embeddings for solving conservative nonlinear ordinary differential equations on quantum computers}
%\title{Dual Koopman--von Neumann embeddings for conservative non-linear ordinary differential equation  quantum solver: numerical limitations }

\author{Thibault Fredon}
\affiliation{\it{Plasma Science and Fusion Center, Massachusetts Institute of Technology, Cambridge, Massachusetts 02139, USA}}
\author{Abhay K. Ram}
\affiliation{\it{Plasma Science and Fusion Center, Massachusetts Institute of Technology, Cambridge, Massachusetts 02139, USA}}
\author{Fabrice Debbasch}
\affiliation{Sorbonne Université, Observatoire de Paris, Université PSL, CNRS, LUX, F-75005 Paris, France}
\author{Julien Zylberman}
\affiliation{CERFACS, 42 Avenue Gaspard Coriolis, 31057 Toulouse Cedex 1, France}
\author{Nuno F. Loureiro}
\affiliation{\it{Plasma Science and Fusion Center, Massachusetts Institute of Technology, Cambridge, Massachusetts 02139, USA}}

\begin{abstract}
The simulation of nonlinear ordinary differential equations on quantum computers is inherently challenging, as quantum gates are linear operators on qubit states. In this paper, we put forth a Koopman-von Neumann (KvN) operator based algorithm for solving nonlinear ordinary differential equations on a quantum computer which overcomes the innate limitations of quantum operations. In this approach, a Liouville probability density is embedded into a wavefunction, the evolution of which is governed by an operator dual to the Koopman operator. The  trajectories corresponding to the nonlinear differential equations are reconstructed from the average value of Koopman observables evaluated through quantum measurements. We specifically evaluate the computational limitations of solving nonlinear equations within this Liouville embedding framework.  
An Ehrenfest-type estimate is derived that relates the reconstruction error to the covariance of the transported density, highlighting the competition between linear stretching and Hessian-induced folding.
This leads to a key stability criterion, which depends on the local Ehrenfest-Reynolds number. We discuss the effects of measurement-induced errors including those due to Hadamard-test sampling, amplitude estimation, and bias due to covariance in probabilistic Grover-type inference on the evaluation of trajectories. The limiting bounds on the accuracy of quantum computations are numerically verified for the Lotka-Volterra system and for the quartic oscillator.  We observe a rapid growth in computational errors when the Ehrenfest-Reynolds number approaches the predicted threshold. The bounds resulting from our analysis provide key guidelines for selecting the width of the initial Gaussian probability distribution associated with a system of nonlinear ordinary differential equations.

\end{abstract}
\maketitle
\section{Introduction}

Solving nonlinear differential equations on quantum computers is a longstanding objective in quantum computing.
The challenge is that quantum algorithms are built from linear operations on quantum states, whereas the dynamics of interest are nonlinear \cite{Leyton09,Liu2021,CLO21,Childs23, Koukoutsis26}.
As a result, one strategy is to construct a linear embedding of the nonlinear system, allowing its dynamics to be implemented as a unitary evolution, either through Hamiltonian simulation \cite{Lloyd1996, Low2017, Low2019, Bastidas2024}, or by using a Quantum Linear Systems Algorithm (QLSA) to solve the resulting linear system \cite{HHL09, Morales24}.
Such embeddings are often problem dependent, as exemplified by the Madelung transform \cite{ZDMD22}, or rely on strong structural assumptions such as Carleman linearization \cite{Liu2021}.
Carleman linearization, in particular, relies on a polynomial basis decomposition of the solution to recast the original non-linear problem into a linear one in an enlarged set of amplitudes \cite{Carleman1932}. While powerful, this approach introduces convergence conditions that restrict its domain of applicability. Moreover, extracting the solution from the resulting quantum state generally requires measurement procedures involving post-selection that significantly increase the associated computational cost.

Alternative approaches based on quantum formalisms have also been developed. These include methods based on second quantization \cite{May2023,may2024,MayPHD}, quantum measurement procedures \cite{Andress2024,Shi2024}, block encodings for non-unitary dynamics \cite{novikau2024, Zylberman2026}, and Completely Positive Trace Preserving (CPTP) maps \cite{Joseph2023}. 
Such approaches provide quantum representations of dynamical systems while relaxing, in some cases, the strict unitarity requirement on the evolution operator at the cost of additional post-processing.

Within this broader effort, the Koopman--von Neumann (KvN) formalism emerged as a promising framework for the treatment of non-linear differential equations. Originally introduced in the early development of quantum mechanics as an operator formulation of classical dynamics \cite{Koopman31,Klein2017}, the KvN framework provides a linear representation of a broad class of dynamical systems and is supported by a substantial operator-theoretic literature \cite{Mezic2005,Budisic2012,Mezic2013,Klus2016,MMS20}.
In the context of quantum computing, KvN-based approaches have been proposed for Liouville equation simulations \cite{Joseph2020, Zylberman2025}, and subsequently extended to dissipative systems through block-encoding techniques \cite{novikau2024,Zylberman2026}.
Within the Koopman operator framework, several generalized linearization procedures have been developed \cite{tanaka2025,novikau2025}.

In this paper, we leverage the duality properties of the Koopman operator to extract the solution of the underlying linear or non-linear ODE system. 
In Sec.~\ref{Sec: KvN}, we present the Koopman operator formalism in general, establishing the connections between nonlinear ODEs, the corresponding Koopman operator, and its dual acting on linear functionals. We then relate the dual Koopman operator to the Liouville equation and show that this formalism can be used to extract the system trajectories.
In Sec.\ \ref{Sec:Transport},  we recall the results of Ref.\cite{Zylberman2025} and establish that the transport equation formalism is connected to the dual Koopman formalism.
In Sec.\ \ref{Sec: ODE Lim}, we show that reconstructing the ODE solution from discretized density averages introduces additional limitations. More precisely, we prove that the accuracy of the reconstructed classical trajectory is primarily controlled by the density variance and by the Hessian of the underlying flow. This allows us to identify an upper bound on the error between the reconstructed trajectory and the actual trajectory.
In Sec.\ \ref{Sec:Measurement}, we define and analyze the measurement step which introduces additional sources of error when reconstructing non-linear trajectories. In particular, we recover the $O(1/\epsilon)$ complexity scaling to achieve an accuracy $\epsilon$.
%In Sec.\ \ref{Sec:Measurement}, we analyze the measurement limitations associated with the dual KvN reconstruction procedure. Since the density-based formulation connects averaged observables to quantum measurement outcomes, it provides a framework suited to near-term quantum implementations which is optimal. However, this measurement step introduces additional sources of error, which must be accounted for when reconstructing non-linear trajectories. 
%
In Sec.\ \ref{Sec: Numerical results}, previous results are supported by the simulation of the Lotka-Volterra system whose non-linearity is well behaved, as well as the quartic oscillator whose phase mixing properties illustrates some limitations due to the non-linearity.

\section{Koopman operator formalism}
\label{Sec: KvN}
\begin{figure}[h]
\centering
\[
\begin{tikzcd}[
  column sep=7em, row sep=6em,
  cells={nodes={anchor=center}},
  >={Stealth[length=6pt]}
]
{\displaystyle
  \begin{aligned}
    \dot{\textbf{x}} = \textbf{F}[\textbf{x}] \\
    S_t(\textbf{x}_0)=\textbf{x}(t)
  \end{aligned}
}
  \arrow[r, "{\text{Observable } f\in \mathcal{C}(\Omega) }"]
  \arrow[d, "{\text{Statistical state}}"']
&
{\displaystyle
  \begin{aligned}
    \dot f = \textbf{F}[\textbf{x}]\!\cdot\!\nabla_{\textbf{x}} f \\
    \mathcal{U}_t\,f(\textbf{x}_0)= f(\textbf{x}(t))
  \end{aligned}
}
  \arrow[d, "{\text{Linear functional}}"]
  \arrow[dl, "{\text{Duality, Riesz + Radon-Nikodym}}" description]
\\
{\displaystyle
  \begin{aligned}
    \partial_t \rho = -\nabla\!\cdot\!(\mathbf{F}\rho) \\
    \mathcal{P}_t\rho(\textbf{x},0)=\rho(\textbf{x},t)
  \end{aligned}
}
  \arrow[r, "{\text{Expectation value}}"]
&
{\displaystyle
  \begin{aligned}
    \int \! d\mathbf{x}\, f(\mathbf{x})\,\rho(\mathbf{x},t) =\\
    \int \! d\mathbf{x}\, f(\textbf{x}(t))\,\rho(\mathbf{x}) 
  \end{aligned}
}
\end{tikzcd}
\]
\caption{Commutative diagram linking the ODE (top left), the observable description (top right), and the density evolution (bottom left) in both a differential and an operator form. Note that even if the Perron Frobenius operator  $\mathcal{P}_t$ is well-defined, its differential representation in terms of Liouville equation is not  guaranteed.}
\label{Fig: Commutative diagram}
\end{figure}
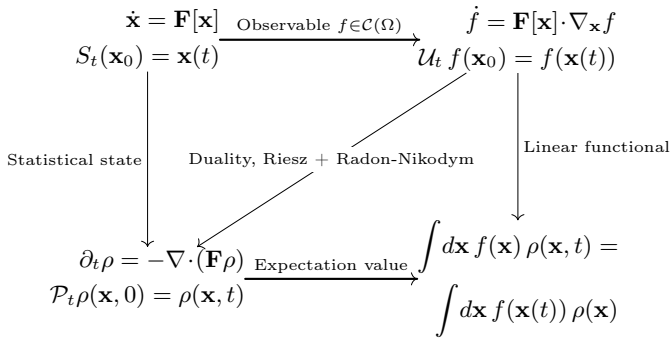

This section establishes a formal connection between measure preserving dynamical systems and the associated unitary Koopman operator representation.

%
% Nevertheless, a general and usefully verifiable criterion ensuring that a dynamical system described by ordinary differential equations admits a unitary Koopman operator remains lacking.
%
%The emphasis is put on the fact that the Koopman operator, which evolves observables, admits a dual representation in terms of the Perron-Frobenius operator, which evolves probability density function \cite{Klus2016,MMS20}. 
%
%The Perron-Frobenius operator in general conserves the norm of the probability density functions which can be represented as a quantum state.
%
%In a volume conserving case, both the Koopman operator and the Perron-Frobenius operator are unitary.
%
%Because this quantum state encodes an entire distribution rather, recovering the solution of a differential equation naturally becomes a problem of measuring suitable observables. 
%
%Finally, the quantum mechanics interpretation is an intrinsic and general feature of the Perron--Frobenius/KvN formalism. 

Consider an autonomous non-linear dynamical system
\begin{equation}
    \frac{d\bf{x}}{dt}= \textbf{F}(\textbf{x}) ,\, \textbf{x}(0) =\textbf{x}_0\in\Omega\subseteq\mathbb{R}^d
    \label{Eq: ODE classic}
\end{equation}
where the function $\textbf{F}(\textbf{x}): \mathbb{R}^d\rightarrow\mathbb{R}^d$ is assumed locally Lipschitz continuous and non-linear, and $\Omega$ is a compact phase space.
The dynamical system can be defined by its \textit{flow}, which is a collection of maps $S_t:\Omega\rightarrow\Omega$ indexed by $t\in \mathbb{R}^+$ verifying $S_t(\textbf{x}_0):=\textbf{x}(t)$.

In this framework, the \textit{observables} of the system are in the Banach space $\mathcal{F}$ of continuous bounded function $f: \Omega\rightarrow\mathbb{R}$.
The evolution of an observable under the flow $S_t$ is represented by the \textit{Koopman Operator}  $\mathcal{U}_t:\mathcal{F}\rightarrow\mathcal{F}$,
\begin{equation}
	\mathcal{U}_t f := f \circ S_t\,.
\end{equation}

For continuous-time dynamics, the Koopman operator is generated by the linear operator
$
\mathcal{L}:=\mathbf{F}(\mathbf{x})\cdot\nabla ,
$
where $\nabla$ denotes the gradient with respect to the phase-space coordinates $\{x_i\}_{i=1}^d$ where $d$ is the dimension of the phase space.

The Koopman operator has a dual operator acting in the functional space $\mathcal{F}^*=\mathcal{C}(\Omega)^*$ \cite{MMS20}.
If $\Xi:\mathcal{F}\mapsto\mathbb{R}$, the duality of Banach space implies that the Koopman operator has a dual representation $\mathcal{U}_t^*: \mathcal{F^*}\rightarrow\mathcal{F^*}$ which acts as $ \mathcal{U}_t^*\Xi(f):=\Xi\left(\mathcal{U}_t f\right) $.

From Riesz theorem\cite{Riesz1909,MMS20}, $\Xi$ can be represented by a measure.
Moreover, if this measure is absolutely continuous, it can be associated to a density function $\rho\in L^1(\Omega)$, following Radon-Nikodym \cite{Radon1913,Nikodym1930,MMS20}.
In turns, $\Xi$ can be written as:
\begin{equation}
	\Xi(f) = \int_\Omega f(\textbf{x})\rho(\textbf{x}) d\textbf{x},
    \label{Eq: Observable average}
\end{equation}
where $\textbf{x}$ a coordinate system that is not necessarily expressed in terms of canonical variables.
For the purpose of quantum algorithms, a linear functional $\Xi$ is an averaging procedure over the probability density $\rho$.

The previous expression for $\Xi$ allows to write the dual of the Koopman operator applied on $\Xi$ as a time evolution of the density $\rho$.
Such an evolution is represented by the \textit{Perron-Frobenius operator} $\mathcal{P}_t: L^1(\Omega)\rightarrow L^1(\Omega)$ \cite{MMS20,Klus2016} as:
\begin{equation}
    \left(\mathcal{U}_{t}^* \Xi\right) (f) = \int_\Omega f(\textbf{x})\,.\,\mathcal{P}_t (\rho)(\textbf{x)} d\textbf{x}
\end{equation}
The previous equation allows one to establish correspondence between the Perron-Frobenius operator and the Koopman operator as 
\begin{equation}
    \int_\Omega (\mathcal{U}_tf)(\textbf{x}) \,\rho (\textbf{x}) d\textbf{x} = \int_\Omega f(\textbf{x})\,(\mathcal{P}_t \rho)(\textbf{x}) d\textbf{x}.
\end{equation} 

The preceding identity provides two dual descriptions of the same dynamics: the Koopman operator pulls observables back along the flow, whereas the Perron--Frobenius operator pushes densities forward, as summarized in Fig.~\ref{Fig: Commutative diagram}.

This duality does not, by itself, imply unitarity. If the flow $S_t$ is invertible and preserves a measure $\mu$, then the Koopman operator is unitary on $L^2(\Omega,\mu)$. In this case, the corresponding Perron--Frobenius operator satisfies
\begin{equation}
    \mathcal{P}_t
    =\mathcal{U}_t^\dagger
    =\mathcal{U}_t^{-1}
    =\mathcal{U}_{-t}.
\end{equation}
For densities defined with respect to the Lebesgue measure, its infinitesimal evolution is governed by the Liouville equation
\begin{equation}
    \partial_t\rho(\mathbf{x},t)
    =-\nabla_{\mathbf{x}}\cdot
    \bigl(\mathbf{F}(\mathbf{x})\rho(\mathbf{x},t)\bigr).
    \label{Eq: Liouv mifa}
\end{equation}

A sufficient condition for the Lebesgue measure to be invariant is that the vector field be divergence-free. More generally, if a change of variables $\mathbf{q}=\mathbf{q}(\mathbf{x})$ renders the transformed vector field divergence-free,
\begin{equation}
\nabla_{\mathbf{q}}\cdot\widetilde{\mathbf{F}}
    =0,
\end{equation}
then the Lebesgue measure in the $\mathbf{q}$ coordinates—and its pullback to the original phase space—is invariant. Canonical Hamiltonian systems provide an important example of this structure, as recalled in App.~\ref{App.: Ham}.
The direct unitary implementation developed in this paper is not applicable when no suitable measure-preserving representation is available such as the one of dissipative systems \cite{LaSalle1972}.
These systems are not excluded from quantum simulation, but require an enlarged or probabilistic construction, such as Schrödingerisation \cite{Liu2023} or block-encoding methods \cite{novikau2025,ZNSD24,Zylberman2026}.

% As a result, the quantum algorithm scheme presented in this paper, though illustrated on specific cases, requires the underlying dynamical system to be measure-preserving in order for the Koopman operator, and thus the Perron–Frobenius operator, to be unitary.
% %
% Dissipative ODEs, as defined in Ref.\ \cite{LaSalle1972}, are not solvable with the present method since their Koopman operator representation is not apriori unitary since dissipative ODE systems presents no measure-preserving structure over the phase space.
% %
% As a consequence, dissipative ODEs dynamics must be embedded into a higher-dimensional unitary evolution since their evolution is not unitary. 
% %
% From a quantum theory perspective, non-unitary dynamics is analogous to decoherence models \cite{Schloss07}. 

\section{Transport equation \& Quantum algorithm}
\label{Sec:Transport}

In Ref.~\cite{Zylberman2025}, Zylberman et al.\ introduced a quantum algorithm for transport equations whose vector field satisfies the component-wise condition
\begin{equation}
    \partial_{x_i}F_i(\mathbf{x})=0,
    \qquad i=1,\ldots,d.
    \label{Eq: Hypothesis}
\end{equation}
Here, $F_i$ denotes the $i$-th component of $\mathbf{F}$.
Since Eq.~\eqref{Eq: Hypothesis} implies $\nabla_{\mathbf{x}}\cdot\mathbf{F}=0$, the Liouville equation reduces to the advection equation
\begin{equation}
    \partial_t\rho(\mathbf{x},t)
    =
    -\mathbf{F}(\mathbf{x})\cdot
    \nabla_{\mathbf{x}}\rho(\mathbf{x},t).
    \label{Eq: Liouville}
\end{equation}

Condition~\eqref{Eq: Hypothesis} is stronger than divergence-freeness and ensures the Hermiticity of each contribution to the transport Hamiltonian, thereby simplifying its quantum implementation by Trotterization. 
It is satisfied by the separable Hamiltonian systems considered in this work, whose Liouville evolution can therefore be simulated using the scheme of Ref.~\cite{Zylberman2025}.

% For measure-preserving systems, the Liouville density is transported along the flow generated by the underlying ODE.
% %
% If the initial Liouville density mean is $\mathbf{x}_0$, the trajectory of its mean provides an approximation of the ODE solution.
% %
% The accuracy of this reconstruction is limited by the finite width of the density, by the discretization of phase space (addressed in Ref.\ \cite{Zylberman2025}), and by the statistical error associated with the measurement procedure. 
% %
% These topics will be discussed below.

In order to develop the ODE solver, two observations need to be considered.
First, by linearity of Eq.\ \ref{Eq: Liouville} and since the function $\rho$ is definite positive, both $\rho$ and $\sqrt{\rho}$ follow the same Liouville equation.
Second, since $\rho$ is non-negative and integrable, $\sqrt{\rho}$ is square-integrable and can be used as a wavefunction amplitude.

To build the discrete quantum state, we consider the Phase Space to be $\Omega = [0,1]^d$.
Each dimension is assumed to be discretized on $n$ qubits, implying a numerical grid of $2^{n\times d}$ points.
Noting $\tilde{\Omega}$ the square grid of size $2^{n\times d}$ representing the discretized phase space, the discretized Koopman wave function is defined as 
\begin{equation}
    \ket{\rho}(t)=\frac{\sum_{\textbf{x}\in\tilde{\Omega}}\sqrt{\rho(\textbf{x},t)}\ket{\textbf{x}}}{\sqrt{\sum_{\textbf{x}\in\tilde{\Omega}}\rho(\textbf{x},t)}}.
\end{equation}
This state can therefore be evolved using the Perron--Frobenius propagator associated with the Liouville Eq.\ \ref{Eq: Liouville}.
Following Ref.\ \cite{Zylberman2025}, the Perron-Frobenius operator can be discretized and decomposed via a Trotter -- Suzuki decomposition \cite{Trotter1959, Suzuki1993}, \textit{i.e.},
\begin{equation}
    \hat{\mathcal{P}}_{T}:= \left(\prod\limits_{i=1}^d e^{-\Delta t \,F_i(\textbf{x})\, \hat{D}_{i,p}}\right)^T,
    \label{Eq: Discretized propagator}
\end{equation}
where $\Delta t$ is the discrete time step, $T$ is the total number of time steps, and $\hat{D}_{i,p}$ is the discrete finite derivative operator along the $x_i$-axis with stencil $p$.
The formal definition of $\hat{D}_{i,p}$ in terms of discrete shift operators $\mathcal{S}$ reads 
\begin{equation}
    \hat{D}_{i,p}:=\frac{\sum_{l=-p}^p a_l\mathcal{\mathcal{S}}^l_{x_i}}{\Delta x_i},
\end{equation}
where $\Delta x_i$ is the grid step size along the $x_i$ axis, $a_0=0$, $a_k=(-1)^{k+1}(p!)^2/(k(p-k)!(p+k)!)$ for $k\in [\![-p,p]\!]\backslash \{0\}$, and the shift operators are defined as ${\mathcal{S}}_{x_i}\ket{x_i}=\ket{x_i+\Delta x_i}$.
Following Ref.\ \cite{Zylberman2025}, the operator can be efficiently implemented in quantum circuits through Quantum Fourier Transform (QFT) and diagonal unitary operators using Walsh decomposition.

ODE trajectories can be computed from the average position of the evolved density.
% % %
% This approach originates from the use of the Quantum Fourier Transform (QFT), which prevents the use of Dirac delta initial conditions \footnote{A Dirac trajectory would theoretically track a trajectory with no covariance induced error. Indeed, distributions whose characteristic width approaches the grid spacing become under-resolved and effectively collapse into grid-scale spikes.}.
%
We still have to account for the covariance matrix magnitude effect, more precisely, the extent to which it introduces systematic errors in the recovered dynamics or measurement procedures.
% %
% To account for these effects, the analysis is shifted from pointwise trajectories to averaged observables (related to the trajectory through linear functionals).
%
% Using the dual formulation of the Koopman formalism, these averaged observables evolution follows an Ehrenfest-type approach of observables (see App.\ \ref{App.: Ham}).
%
In the next section, the errors induced by the covariance are quantified.

\section{ODE solving and limits}
\label{Sec: ODE Lim}

The ODE solution can be reconstructed on a quantum computer by measuring the average position of the probability density in phase space.
The coordinate functions $x_i: \Omega \rightarrow \mathbb{R}$ are observables of the system.
The objective is to match the prediction of the linear functional

\[
\langle x_i\rangle_\rho:=\Xi_\rho[x_i]
=
\int_\Omega x_i\,\rho(\mathbf{x},t)\,d\mathbf{x},
\]
with the exact solution of the ODE, denoted by $\tilde{x}_i$.
For nonlinear vector fields $\textbf{F}$, the evolution of this mean depends on higher moments of the density and therefore generally deviates from the exact ODE trajectory. In this section, we bound this discrepancy in terms of the variance of the evolving density.

In the Perron-Frobenius picture, the observable $x_i$  are static in time while the probability density evolves, the dynamics of $\langle x_i\rangle_\rho$ are governed by an Ehrenfest-type relation (App.\ \ref{App.: Ham}),
\begin{equation}
\frac{d\langle x_i\rangle_\rho}{dt}
=
\langle\textbf{F}[\textbf{x}]\!\cdot\!\nabla_{\textbf{x}} x_i\rangle_\rho
=
\langle F_i(\textbf{x})\rangle_\rho .
\label{Eq: Ehr Eq}
\end{equation}

\subsection{Lipschitz bound}

Assuming that $F_i$ is $K_i$-Lipschitz, the growth rate of the error
$\langle x_i\rangle_\rho-\tilde{x}_i$
is bounded by
\begin{equation}
\left|
\frac{d(\langle x_i\rangle_\rho-\tilde{x}_i)}{dt}
\right|
\leq
K_i
\sqrt{
\sum_{j=1}^d \operatorname{Var}_{\rho(t)}(x_j)
+
\|\langle \textbf{x}\rangle_\rho-\tilde{\textbf{x}}(t)\|^2
},
\label{Eq: Ehrenfest bound}
\end{equation}
where
\[
\|x\|
=
\sqrt{\sum_{i=1}^d |x_i|^2}
=
\max_{\boldsymbol{\zeta}\in\mathbb{S}^{d-1}}
\sum_{i=1}^d \zeta_i x_i ,
\]
and $\mathbb{S}^{d-1}$ denotes the unit sphere of $\mathbb{R}^d$ namely $\mathbb{S}^{d-1}=\{\mathbf{x}\in \mathbb{R}^d: \|\mathbf{x}\|=1\}$.
A detailed derivation is provided in App.\ \ref{App: Ehr bound}.
The bound \ref{Eq: Ehrenfest bound}  motivates the introduction of quantities characterizing both the drift of the averaged trajectory and the spatial spread of the probability density. We respectively define the \emph{error vector}, the \emph{covariance matrix}, and the associated \emph{empirical covariance matrix} as
\begin{align}
\epsilon_i
&:=
\langle x_i\rangle_{\rho(t)}-\tilde{x}_i(t),
\label{Eq: err vect}
\\
\Sigma_{ij}(t)
&:=
\langle
(x_i-\langle x_i\rangle_\rho)
(x_j-\langle x_j\rangle_\rho)
\rangle_\rho ,
\\
\tilde{\Sigma}_{ij}(t)
&:=
\langle
(x_i-\tilde{x}_i(t))
(x_j-\tilde{x}_j(t))
\rangle_\rho
\nonumber
\\
&=
\Sigma_{ij}(t)
+
\epsilon_i(t)\epsilon_j(t).
\label{Eq: Emp Cov}
\end{align}
The term on the right hand side of \ref{Eq: Ehrenfest bound} can be expressed as, $
\sqrt{\operatorname{Tr}(\tilde{\boldsymbol{\Sigma}})}
=
\|\tilde{\boldsymbol{\Sigma}}^{1/2}\|_F $.
Here, $\|\cdot\|_F$ denotes the Frobenius norm, defined for
$A\in\mathbb{R}^{d\times d}$ by
\begin{equation}
\|A\|_F
:=
\left(\sum_{i,j=1}^{d}|A_{ij}|^2\right)^{1/2}
=
\sqrt{\operatorname{Tr}(A^{\top}A)}.
\label{Eq: Frob Norm}
\end{equation}
Thus, the bound \ref{Eq: Ehrenfest bound}  controls the growth of the trajectory error through the root-mean-square spread of the density around the target trajectory.
The choice of the Frobenius norm is motivated by the fact that it captures an average root-mean-square contribution over all dimensions, whereas the canonical matrix two norm $\|A\|=\max_{\boldsymbol{\zeta}\in\mathbb{S}^{d-1}}\|A\zeta\|$ is controlled by the largest directional contribution.

The bound \ref{Eq: Ehrenfest bound} already highlights two important requirements :
\begin{enumerate}
\item the probability density should remain centered around the target solution;
\item the probability density should remain sufficiently localized, i.e.\ its empirical covariance matrix must remain small.
\end{enumerate}

\subsection{Taylor-expansion error bound under constant-covariance closure}

We illustrate the bound in \ref{Eq: Ehrenfest bound} for an analytical flow in phase space $\Omega$. Using index sum convention over indices $j$ and $k$, a Taylor series expansion leads to :
\begin{align}
    \langle F_i(\textbf{x})\rangle_\rho \simeq F_i(\tilde{\textbf{x}})+\epsilon_j\,\partial_{x_j}F_i|_{\tilde{\textbf{x}}}+\frac{\tilde{\Sigma}_{jk}}{2}\partial_{x_j}\partial_{x_k}F_i|_{\tilde{\textbf{x}}}
    \nonumber
\end{align}

As the Lipschitz bound \ref{Eq: Ehrenfest bound}, the previous expression accentuates the two types of error sources.
The first \textit{trajectory drift term} coming from linear separation between the two trajectories $\langle x_i\rangle$ and $\tilde{x}_i$.
The second order term is related to the \textit{empirical covariance}.
The bound therefore captures the competition between the amplification of the trajectory mismatch, governed by the Jacobian of the flow, and the correction induced by phase-space spreading, governed by its Hessian.
The truncation at second order nevertheless constitutes a closure assumption: for an arbitrary distribution, higher-order statistical contributions need not be small.
Retaining all orders would instead produce a hierarchy of statistical moments, analogous to a Carleman expansion \cite{Carleman1932}, which would itself require closure.

Here, we close the hierarchy at second order by assuming that the densities remain approximately Gaussian, so that cumulants of order three and higher are negligible and the remaining moments are determined by the mean and covariance.
Under these closure assumptions, we obtain

\begin{equation}
\frac{d\epsilon_i}{dt}
=
\epsilon_j \,\partial_{x_j}F_i|_{\tilde{\mathbf{x}}}
+
\frac{\Sigma_{jk}}{2}
\partial_{x_j}\partial_{x_k}F_i|_{\tilde{\mathbf{x}}}
+
O(\|\boldsymbol{\epsilon}\|^2,\|\boldsymbol{\Sigma}\|^2).
\end{equation}

To estimate the error growth, we use the Frobenius norm as defined in Eq.\ \ref{Eq: Frob Norm}.
This norm can be adapted to the covariance-Hessian contraction since
\[
\left|
\Sigma_{jk}\partial_{x_j}\partial_{x_k}F_i|_{\tilde{\mathbf{x}}}
\right|
=
\left|
\operatorname{Tr}\!\left(
\Sigma^{\top}D^2F_i|_{\tilde{\mathbf{x}}}
\right)
\right|
\leq
\|\Sigma\|_F
\,
\|D^2F_i|_{\tilde{\mathbf{x}}}\|_F .
\]
where $D^2F_i$ denotes the Hessian matrix with entries
$(D^2F_i)_{jk}=\partial_{x_j}\partial_{x_k}F_i$.

Moreover,
\[
\left|
\epsilon_j \partial_{x_j}F_i|_{\tilde{\mathbf{x}}}
\right|
\leq
\|\boldsymbol{\epsilon}\|
\,
\|\nabla F_i|_{\tilde{\mathbf{x}}}\| .
\]
Therefore,

\begin{equation}
\left|
\frac{d\epsilon_i}{dt}
\right|
\leq
\|\boldsymbol{\epsilon}\|
\,
\|\nabla F_i|_{\tilde{\mathbf{x}}}\|
+
\frac{1}{2}
\|\boldsymbol{\Sigma}\|_F
\,
\|D^2F_i|_{\tilde{\mathbf{x}}}\|_F
+
O(\|\boldsymbol{\epsilon}\|^2,\|\boldsymbol{\Sigma}\|^2),
\end{equation}

Although maximizing over the phase space $\Omega$ removes the explicit dependence on $\widetilde{\mathbf{x}}$, the resulting differential inequality cannot be integrated without a closure assumption for the covariance matrix $\boldsymbol{\Sigma}$.
For the short-time regime considered here, we impose
\[
\left\lVert\boldsymbol{\Sigma}(t)\right\rVert_F
=
\left\lVert\boldsymbol{\Sigma}(t_0)\right\rVert_F.
\]
This assumption is substantially stronger than phase-space volume conservation: a volume-preserving flow may deform the distribution through stretching and contraction, whereas the present closure suppresses any variation of its overall covariance scale.
Despite its restrictive nature, it provides a tractable first approximation and leads to an explicit criterion for selecting the initial covariance.

To express the resulting bound compactly, we introduce
\begin{align}
\epsilon(t)
&:=
\left\lVert\boldsymbol{\epsilon}(t)\right\rVert_2
\leq
\sqrt{d}\,
\max_{1\leq i\leq d}|\epsilon_i(t)|,
\notag\\
L
&:=\sqrt{d}
\max_{1\leq i\leq d}
\sup_{\mathbf{x}\in\Omega}
\left\lVert\nabla F_i(\mathbf{x})\right\rVert_2,\\
H
&:=\sqrt{d}
\max_{1\leq i\leq d}
\sup_{\mathbf{x}\in\Omega}
\left\lVert D^2 F_i(\mathbf{x})\right\rVert_F,
\notag\\
\sigma_F^2(t)
&:=
\frac{
\left\lVert\boldsymbol{\Sigma}(t)\right\rVert_F
}{
\sqrt{d}
}.
\label{Eq: Frob sigma}
\end{align}
This normalization ensures that $\sigma_F=\sigma$ for an isotropic covariance matrix $\boldsymbol{\Sigma}=\sigma^2I_d$.

Under the constant-covariance approximation, integration of the differential inequality yields
\begin{align}
0\leq\epsilon(t)
\leq{}&
e^{L(t-t_0)}\,\epsilon_0
+
\sigma_F(t_0)\,R
\left(
e^{L(t-t_0)}-1
\right),
\label{Eq: Ehrenfest Bound 1st order}
\end{align}
where $\epsilon_0:=\epsilon(t_0)$ and is determined by the initialization.
The calculation naturally identifies the dimensionless parameter
\begin{equation}
R:=
\frac{
\sqrt{d}\,\sigma_F(t_0) H
}{
2L
}
,
\label{Eq: Local Ehrenfest Reynolds}
\end{equation}
which we call the \emph{Ehrenfest Reynolds number}.
Rather than being introduced independently, $R$ emerges directly from the computation and characterizes the competition between the amplification of the initial trajectory mismatch and the covariance-induced correction.

\subsection{Numerical criterion under constant covariance closure}

Eq.\ \ref{Eq: Ehrenfest Bound 1st order} suggests choosing $\epsilon_0$ and
$\sigma_F \,R$
to be of the same order of magnitude, since both contributions grow at the same rate.
Expressed in terms of the effective covariance width, the numerical criterion becomes
\begin{equation}
\frac{\sigma_F}{\epsilon_0}\sim\frac{1}{R}
\label{Eq: Numerical criterion}
\end{equation}
A numerical choice for $\epsilon_0$ is the grid spacing, since its initial value is assumed to be a purely discretization induced error.
With the current discretization, this corresponds to
$
\epsilon_0\sim\Delta x=\frac{1}{2^{n}}
$
for a simulation involving a total of $n$ qubits per dimension.

Consequently, the ratio $\epsilon_0/\sigma_F$ can be interpreted as the characteristic number of grid points over which the Gaussian initial condition spreads.
To avoid the formation of effectively under-resolved structures, the Gaussian width must remain sufficiently resolved by the numerical grid.
This requires
$\frac{\sigma_F}{\epsilon_0}\gtrsim 10,$
which in turn implies
\begin{equation}
R\lesssim 10^{-1} \label{Eq: Stability criterion}.
\end{equation}

with the choice of $10$ being an arbitrary but reasonable order of magnitude estimate.
This provides an apriori numerical criterion for the validity of the reconstruction scheme, independently of the details of the discretization.

\subsection{Taylor-expansion error bound without constant-covariance closure}

One could also lift the condition constant-covariance approximation by naively bounding $\boldsymbol{\Sigma}(t)$ as
\[
\|\tilde{\boldsymbol{\Sigma}}(t)\|_F
\leq
\|\tilde{\boldsymbol{\Sigma}}(t_0)\|_F
e^{2L|t-t_0|},
\]
The detailed derivation is presented in Subsec.\ \ref{Subapp: Reynolds bound}.
Expressed in terms of \(R\), the bound \ref{Eq: Ehrenfest Bound 1st order} becomes
\begin{equation}
0 \leq \epsilon(t)
\leq
e^{L(t-t_0)}\,\epsilon_0
+
\,
\sigma_F(t_0)\,
R(t_0)
\left(
e^{2L(t-t_0)}-1
\right).
\end{equation}
Using the same reasoning as before, a sufficient asymptotic condition for the reconstruction scheme to remain accurate is

\begin{equation}
R(t_0)\lesssim \frac{e^{-2L|t-t_0|}}{10}.
\end{equation}
The previous criterion is a crude upper-bound as its derivation doesn't involve the volume conservation property of the dynamical system which avoids to ensure the covariance magnitude conservation.

\subsection{Geometrical interpretation}

\begin{figure}
    \centering
    \includegraphics[width=\linewidth]{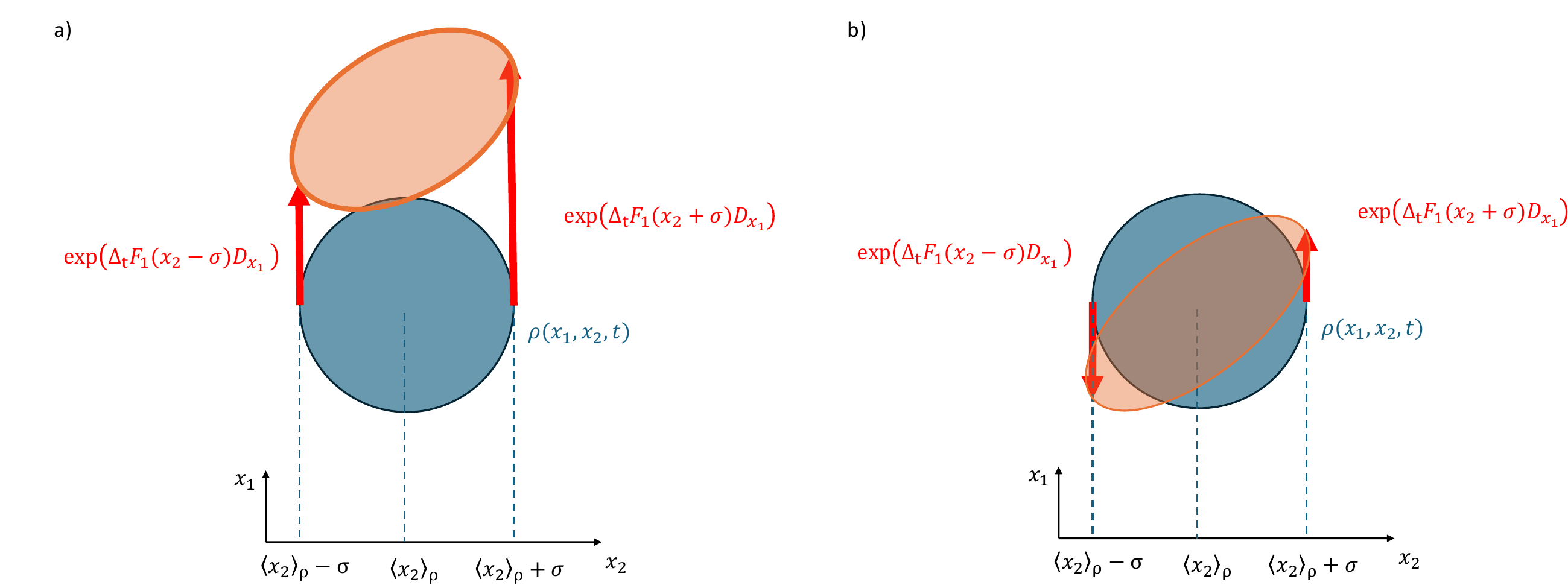}
    \caption{2D representation of the density $\rho(x_1,x_2)$ Full Width at Half Maximum evolution under one Trotter step along  an axis. In Figure 2a), we can see a typical flow from the approach. In 2b), the initial gaussian is assumed to cover a stationary point where $F_i=0$ and to change sign around it.  Volume is conserved as a consequence of Liouville theorem. Variance is contracted along one axis and spread on the other. }
    \label{Fig: Gaussian shear}
\end{figure}
Note that although this criterion was originally derived from the continuous Ehrenfest analysis, it does admit a purely geometric interpretation.
According to Eq.~\ref{Eq: Discretized propagator}, the evolution of the distribution after an intermediate Trotter step can be represented geometrically as illustrated in Fig.~\ref{Fig: Gaussian shear}.
The bound \ref{Eq: Ehrenfest Bound 1st order} can be derived by a geometrical analysis of neighboring trajectories.
For a given effective covariance width $\sigma_F$, we introduce the maximal relative drift

\begin{align}
\Delta_{\sigma_F} x_i
:&=
\max_{\boldsymbol\xi\in \mathcal{B}^d}
\left|
x_i(\textbf{x}+\sigma_F\boldsymbol{\xi},\Delta t)
-
x_i(\textbf{x},\Delta t)
\right|
\\
&=
\Delta t
\max_{\boldsymbol\xi \in \mathcal{B}^d}
\left|
F_i(\textbf{x}+\sigma_F\boldsymbol\xi)
-
F_i(\textbf{x})
\right| +O(\Delta t^2).
\nonumber
\end{align}

Here, $\mathcal{B}^d$ denotes unit ball in dimension $d$.
As shown in App.~\ref{App: Non-Phase Mixing}, the geometrical version of criterion \ref{Eq: Numerical criterion} is

\begin{equation}
\frac{
\sqrt{\sum_i(\Delta_{\sigma_F}x_i)^2}
}{
\sqrt{d}\,\sigma_F
}
\lesssim
10^{-1}.
\label{Eq: Pract crit}
\end{equation}

\section{Quantum measurement}
\label{Sec:Measurement}

In the previous section, the absolute error associated with the dynamical reconstruction scheme was shown to be controlled by the covariance matrix.
Since the output of the quantum algorithm is obtained through measurement, additional uncertainties arising from the estimation of averaged observables must also be taken into account.
%
%In Ref.~\cite{Zylberman2025}, the measurement of the observables introduced in Sec.~\ref{Sec: KvN} is performed using either a Hadamard test or a SWAP test as suggested in Ref.\ \cite{Aaronson2020}.
%
%While the SWAP test can, in principle, provide improved statistical accuracy, it requires a duplication of the quantum registers and therefore doubles the number of qubits involved in the computation.
%
%More generally, the estimation of expectation values on quantum computers has been extensively studied, and several alternative measurement strategies have been proposed \cite{Rall2020,Rall2021}. These approaches may offer substantial reductions in measurement cost and could potentially improve the efficiency of the present algorithm. However, assessing their performance in the context of KvN simulations requires a dedicated analysis and is left for future work.
%

\begin{figure}
    \centering
    \begin{quantikz}[row sep=0.45cm, column sep=0.4cm]
    \lstick[wires=2]{$\ket{\rho}(t)$}
    & \qwbundle{n}
    & \qw
    & \gate[wires=2]{e^{i\theta(X_1,X_2)}}
    & \qw
    & \qw
    \\
    & \qwbundle{n}
    & \qw
    & \qw
    & \qw
    & \qw
    \\
    \lstick{$\ket{0}$}
    & \qw
    & \gate{H}
    & \ctrl{-2}
    & \gate{H}
    & \meter{}
\end{quantikz}
    \caption{Measurement part of the quantum algorithm for a 2-D system where State Prep is a state preparation procedure, $\mathcal{P}_t$ is the unitary operator defined in \ref{Eq: Discretized propagator}, $H$ is the standard Hadamard gate, $e^{i\theta}$ is a diagonal unitary gate. The controlled diagonal unitary operator can be implemented through a Walsh decomposition as explained in App.~\ref{App: Walsh}.}
    \label{fig: Measurement algo}
\end{figure}
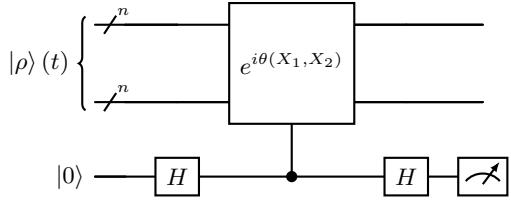

In the following, we focus on two versions of the Hadamard test, a first one achieving a $O(\epsilon^{-2})$ scaling and a second one achieving a $O(\epsilon^{-1})$ scaling using an amplitude amplification technique.
Both methods can be implemented by the circuit shown in Fig.~\ref{fig: Measurement algo} for two different observables $\theta$.

Let $f:\Omega\rightarrow[0,1]$ be an observable of interest. In the Hadamard-test circuit shown in
Fig.~\ref{fig: Measurement algo}, choosing the diagonal phase as
\begin{equation}
\theta(X_1,X_2)
=
2\arccos\sqrt{f(X_1,X_2)}
\end{equation}
yields the following probability of measuring the ancilla in the state
$\ket{0}$:
\begin{align}
    \mathbb{P}_0(t)&=\frac{1}{\mathcal{N}}\sum_{(i_k)\in [\![1,2^{n}]\!]^d}f(x_{i_1},\cdots,x_{i_d})\,\rho(x_{i_1},\cdots,x_{i_d},t)\\&=\int_\Omega f(\mathbf{x})\,\rho(\mathbf{x},t)\, d^d\mathbf{x} +O(2^{-n})
    \label{Eq: P0 eps2}
\end{align}
where $\rho(\mathbf{x},t)$ is the  probability density and $\mathcal{N}$ is accounting for the normalization of $\rho$ on the grid.

Furthermore, assuming that $\rho$ is a multivariate Gaussian distribution with mean $\langle\mathbf{x}\rangle$ and covariance matrix $\mathbf{\Sigma}$, and that $f$ is twice continuously differentiable in a neighborhood of $\langle\mathbf{x}\rangle$, Eq.\ \ref{Eq: P0 eps2} second-order approximation is

\begin{equation}
\int_\Omega f(\mathbf{x})\,\rho(\mathbf{x},t)\, d^d\mathbf{x}
=
f(\langle\mathbf{x}\rangle)
+
\frac{1}{2}
\operatorname{Tr}
\left(
\mathbf{\Sigma}
(D^2f)(\langle\mathbf{x}\rangle)
\right)
+
O(\|\mathbf{\Sigma}\|^2).
\label{Eq: measurement formula}
\end{equation}
As a result, we can directly measure not only the system mean trajectory but also its covariance through this measurement protocol.
Note that, this calculation may hold for other densities providing that higher centered moments $\langle (X-\langle X\rangle_\rho)^k\rangle_\rho$ are negligible with respect to the length scale given by the covariance.
Moreover, Eq.\ \ref{Eq: measurement formula} yields to a close form if $f=x_i$ or $f=x_i^2$, implying that the exact variance and mean can be inferred with this protocol.
With this approach, estimating $\mathbb{P}_0$ with precision $\epsilon$ requires a number of measurement $M=O\left(\frac{\Sigma}{\epsilon^2}\right)$ as detailed in App.~\ref{App.: Bayes}.

Quantum amplitude amplification \cite{Brassard2002,Rall2020, Rall2021} can reduce the number of circuit evaluations
required to achieve a precision $\epsilon$ from
$O(\epsilon^{-2})$ to $O(\epsilon^{-1})$. To apply this approach to the
observable $f$, we encode it in the diagonal phase by setting
\begin{equation}
\theta(X_1,X_2)=f(X_1,X_2)-\theta_m.
\end{equation}
where $\theta_m$ is a reference phase defined by the user.
The expectation value $\langle f\rangle_\rho$ is then inferred through $\theta_m$ by a Grover-like approach \cite{Grover96,Brassard2002} that maximizes 
\begin{equation}
\mathbb{P}_0(\theta_m)
=
\left\langle
\cos\!\left(2(f-\theta_m)\right)
\right\rangle_\rho .
\end{equation}

For a sufficiently localized density, an expansion around
$\langle f\rangle_\rho$ gives
\begin{equation}
\mathbb{P}_0(\theta_m)
=
\cos\!\left(2(\langle f\rangle_\rho-\theta_m)\right)
+
O\!\left(\operatorname{Var}_\rho(f)\right),
\end{equation}
where
\begin{equation}
\operatorname{Var}_\rho(f)
=
\left\langle
\left(f-\langle f\rangle_\rho\right)^2
\right\rangle_\rho .
\end{equation}
For a smooth observable and a localized distribution with mean
$\langle\mathbf{x}\rangle_\rho$, this variance satisfies, at
leading order,
\begin{equation}
\operatorname{Var}_\rho(f)
=
\nabla f(\langle\mathbf{x}\rangle_\rho)^{T}
\boldsymbol{\Sigma}
\nabla f(\langle\mathbf{x}\rangle_\rho)
+
O\!\left(\|\boldsymbol{\Sigma}\|^{3/2}\right).
\end{equation}
A detailed derivation is provided in App.~\ref{App.: Bayes}.
The inferred phase $\theta_m$ is therefore affected by both the statistical
error of amplitude estimation and corrections arising from the finite spread
of the transported density although we reduced the sampling complexity to $O(\epsilon^{-1})$.

Taken together, these results distinguish two limitations of the readout. The Hadamard-test protocol directly estimates averaged observables and, for $f=x_i$ and $f=x_i^2$, provides the corresponding mean and variance required to monitor the stability criterion. 
For a general smooth observable, however, the finite covariance of the transported density controls the difference between the measured average and the corresponding pointwise classical quantity.

To optimize the measurement process of an observable coordinate, i.e., of the reconstructed solution of the nonlinear ODE, one can balance the statistical measurement error with the intrinsic localization scale by choosing
\begin{equation}
\Delta \theta_m=\epsilon \sim \sqrt{\|\mathbf{\Sigma^{1/2}}\|_F}=\sigma_F.
\end{equation}
where $\Delta\theta_m$ is the precision in the measurement of $\theta_m$.
Requesting a substantially smaller measurement error would increase the measurement cost without improving the overall reconstruction accuracy. Combined with Criterion~\ref{Eq: Pract crit} and the requirement that the initial density be resolved by the computational grid, this condition constrains the grid spacing and, consequently, the number of qubits required per dimension. It thereby complements the resource analysis of Ref.~\cite{Zylberman2025}.

\section{Numerical results}
\label{Sec: Numerical results}
\begin{figure*}
\centering

\includegraphics[width=0.45\linewidth]{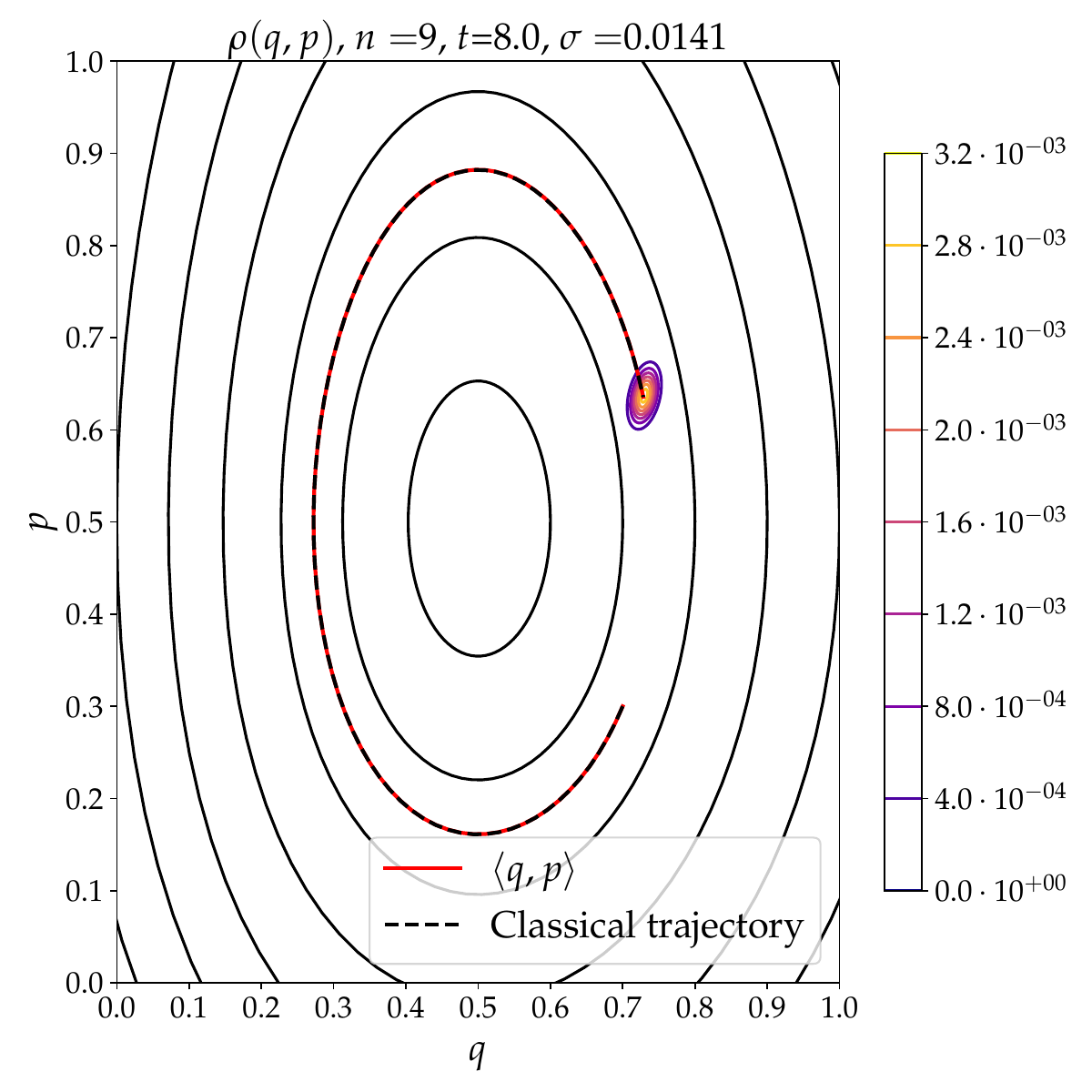}
\hfill
\includegraphics[width=0.45\linewidth]{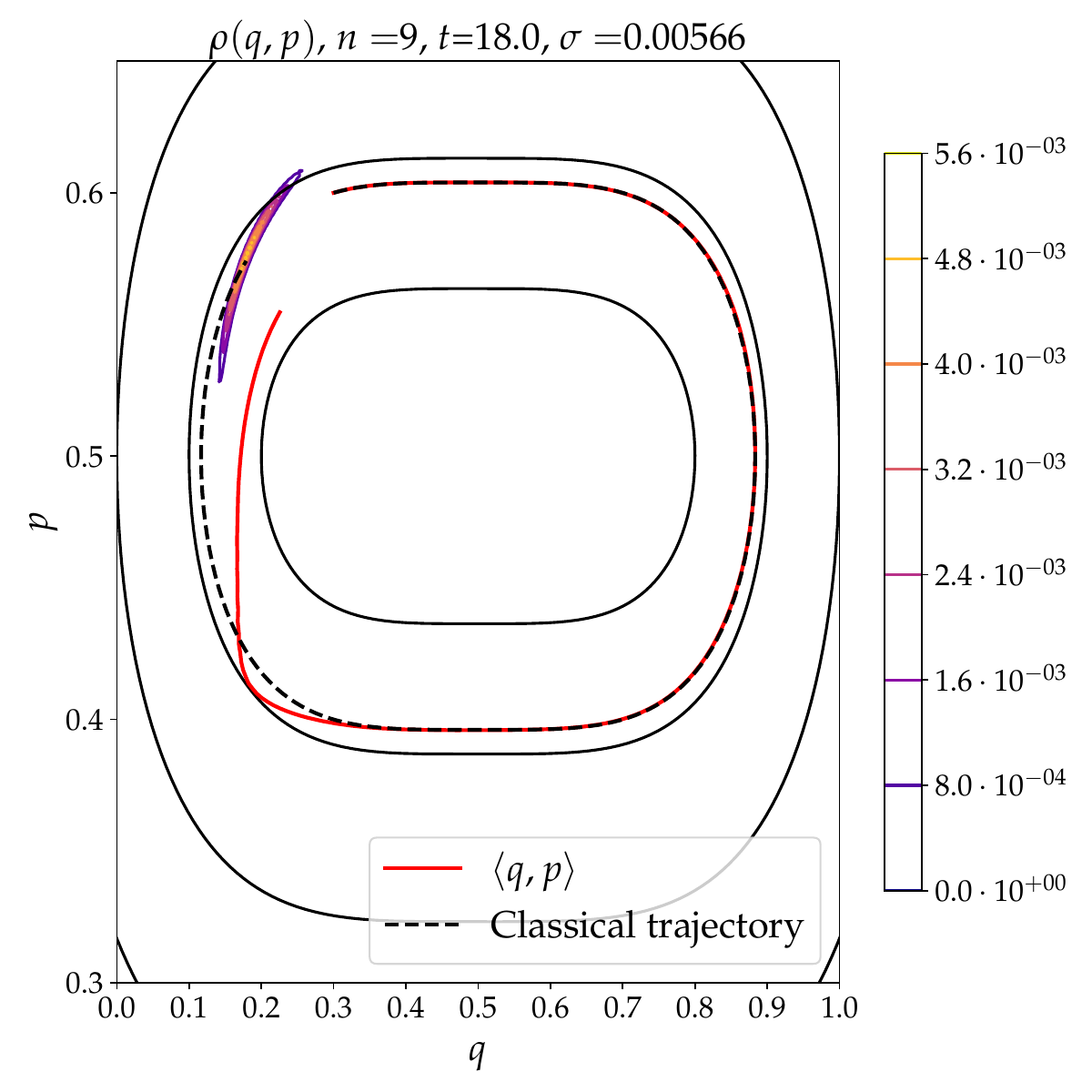}

\caption{Snapshots of the Liouville density for the quartic oscillator with initial gaussian size $\sigma=2\sqrt{8}\times 10^{-2}$ (left) and the Lotka--Volterra system for an initial Gaussian size $\sigma=\sqrt{2}\times 10^{-2}$ (right) with the trace of the mean compared with the classical trajectory computed with a RK4 scheme. The quartic oscillator exhibits strong phase mixing around $p=0$, leading to a rapid deformation of the probability density, whereas the Lotka--Volterra dynamics remain comparatively localized around the stable orbit. Both simulations were performed using the discretization scheme described in Sec.\ \ref{Sec:Transport}.
}

\label{fig:Quartic_LV_snapshots}
\end{figure*}

To illustrate the previous bound and the effect of the covariance, we chose two simple non-linear systems keeping both the dimensionality and the visualization simple.
As a first numerical benchmark, we consider the Lotka--Volterra model, a standard nonlinear system describing the interaction between a prey population $x_1$ and a predator population $x_2$:
\begin{subequations}
\label{Eq: Lokta}
\begin{align}
    \frac{dx_1}{dt} 
    &= \alpha x_1-\beta x_1x_2,\\
    \frac{dx_2}{dt}
    &= -\delta x_2+\gamma x_1x_2.
\end{align}
\end{subequations}
Here, $\alpha$ and $\delta$ are respectively the prey growth and predator mortality rates, while $\beta$ and $\gamma$ characterize their interaction. For positive parameters, the system exhibits periodic trajectories around the coexistence equilibrium
\begin{equation}
    (x_1^\ast,x_2^\ast)
    =
    \left(\frac{\delta}{\gamma},\frac{\alpha}{\beta}\right).
\end{equation}

Introducing the logarithmic coordinates
$(q,p)=(-\ln x_1,-\ln x_2)$, as detailed in
App.~\ref{Ap: LV}, gives the Liouville equation
\begin{equation}
    \partial_t\rho(q,p,t)
    =
    \left(\alpha-\beta e^{-p}\right)\partial_q\rho
    -
    \left(\delta-\gamma e^{-q}\right)\partial_p\rho .
\end{equation}
For the numerical simulation presented in Fig.\ \ref{fig:Quartic_LV_snapshots}, we choose
$\alpha=e^{-1/2-1/3}$, $\beta=e^{-1/3}$,
$\delta=1$, and $\gamma=e^{1/2}$. With this choice, the coexistence equilibrium is located at
$(x_1^\ast,x_2^\ast)=(e^{-1/2},e^{-1/2})$, corresponding to
$(q^\ast,p^\ast)=(1/2,1/2)$. The resulting periodic dynamics provide a simple benchmark for evaluating the reconstruction bound while keeping the transported density comparatively localized.
For the Lotka--Volterra system, the Ehrenfest Reynolds number scales with the initial covariance width as
\begin{equation}
    \frac{R}{\sigma}
    =
    \frac{\gamma}{\beta}
    =
    2.29.
\end{equation}
Consequently, the condition $R<1$ requires
\[
    \sigma<\frac{\beta}{\gamma}\simeq 0.437.
\]
An initial width $\sigma\sim10^{-2}$ therefore gives $R\simeq0.0229$ and comfortably satisfies the criterion. The initial condition used in Fig.~\ref{fig:Quartic_LV_snapshots} is thus expected to provide an accurate reconstruction of the reference trajectory as verified.
% %
% Note that the original Lotka Volterra system presents a diverging Carleman linearized according to Ref.\ \cite{Liu2021} however if the system is centered around its stable attractor the precision drastically \cite{novikau2025}.

As a second and more demanding benchmark, we consider the quartic oscillator. Its Hamiltonian is
\begin{equation}
    H(q,p)=\frac{(p-p_0)^2}{2}+\frac{(q-q_0^4}{4},
\end{equation}
where $q_0=p_0=1/2$ and the corresponding Hamilton equations are
\begin{subequations}
\label{Eq: quartic oscillator}
\begin{align}
    \frac{dq}{dt} &= p-p_0,\\
    \frac{dp}{dt} &= -(q-q_0)^3.
\end{align}
\end{subequations}
The associated Liouville equation is therefore
\begin{equation}
    \partial_t\rho(q,p,t)
    =
    (q-q_0)^3\partial_p\rho
    -
    (p-p_0)\partial_q\rho .
\end{equation}
Because the oscillation frequency depends on the energy, an initially localized density progressively undergoes phase mixing. This deformation is particularly pronounced near the turning points, where $p=p_0$, making the quartic oscillator a stringent test of the covariance-based reconstruction criterion.

Using the maximal value of $\lvert\langle q\rangle\rvert$ reached along the trajectory to derive the Ehrenfest Reynolds, Criterion~\ref{Eq: Stability criterion} predicts an admissible initial width of approximately
\begin{equation}
    \sigma \lesssim 10^{-2}.
\end{equation}
We therefore choose $\sigma=5\times10^{-3}$, which lies within the predicted stability regime.

Both systems are simulated using the discretization scheme of Ref.~\cite{Zylberman2025}, with a sixth-order centered finite-difference stencil, a time step $\Delta t=10^{-3}$, and $n=9$ qubits per dimension. The corresponding phase-space resolution is
$\Delta q=\Delta p=2^{-9}$.

Figure~\ref{fig:Quartic_LV_snapshots} presents snapshots of the transported densities and compares their reconstructed mean trajectories with RK4 exact solutions. For the Lotka--Volterra system, the density remains localized and its mean accurately follows the reference trajectory, with errors below the grid resolution. By contrast, the quartic-oscillator density develops pronounced filamentation, causing the reconstructed mean to progressively depart from the exact trajectory.
%%%
\begin{figure*}[t]
    \centering
    \includegraphics[
        width=0.90\textwidth
    ]{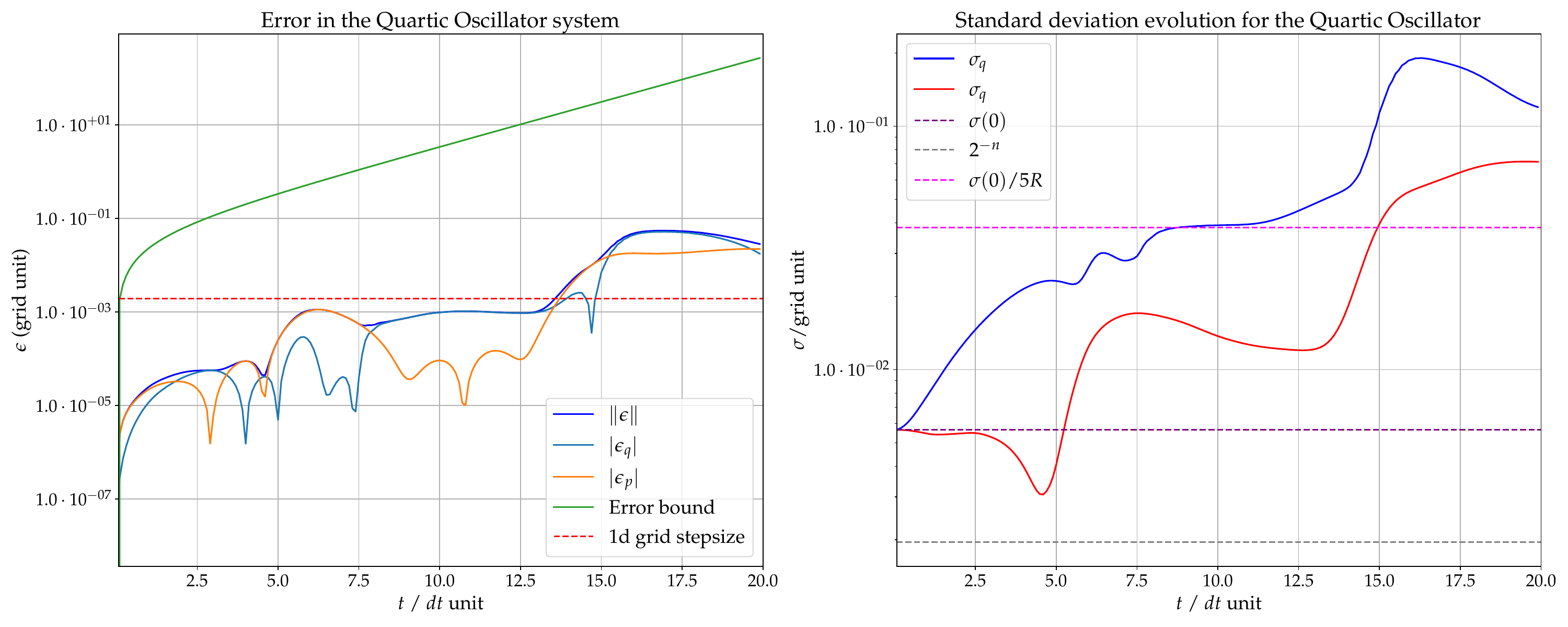}
\caption{
Semi-logarithmic evolution of the reconstruction error and covariance components for the quartic oscillator. Left: error between the RK4 reference trajectory and the mean trajectory reconstructed from the Liouville density, together with the error bound derived in Eq.~\ref{Eq: Ehrenfest Bound 1st order}. The bound largely overestimates the absolute error, but correctly captures the onset and trend of the error growth. Right: evolution of the covariance components. One variance component crosses the threshold associated with the local Ehrenfest Reynolds criterion at approximately the same time as the rapid increase in reconstruction error, supporting the interpretation that phase-space spreading controls the breakdown of the trajectory reconstruction.
}
\label{fig:Quartic_error_variance}
\end{figure*}

Figure~\ref{fig:Quartic_error_variance} shows the time evolution of the reconstruction error and of the covariance components in semi-logarithmic scale.
The bound in Eq.~\ref{Eq: Ehrenfest Bound 1st order} correctly captures the growth trend and provides a key indicator when the computed trajectory is no longer close to the actual solution. Its magnitude significantly overestimate the observed error because of the approximations introduced in its derivation.
Moreover, the increase in both the covariance and the reconstruction error occurs when the transported density approaches the region near $p=0$. Along the quartic oscillator trajectory, this region corresponds to large values of $|q|$, where the Hessian of the vector field becomes large. We have
\[
D^2F_q=0,
\qquad
D^2F_p=
\begin{pmatrix}
-6(q-q_0) & 0\\
0 & 0
\end{pmatrix}.
\]
Thus, the nonlinearity is strongest the curved part of the trajectory in Fig.\ \ref{fig:Quartic_LV_snapshots}.
This behavior is consistent with the analysis developed in Sec.~\ref{Sec: ODE Lim}. The Jacobian controls the stretching of the Gaussian density, while the Hessian controls its nonlinear folding. In the quartic oscillator case, this Hessian-induced folding is the dominant mechanism responsible for the observed reconstruction error.

\section{Conclusion \& Discussion}

Solving non-linear ODEs on quantum computers is particularly challenging, since primitive quantum operations are linear. To address this challenge, we have developed a quantum numerical solver which recovers non-linear trajectories from averaged observables and assess its limitations.

The method was formulated through the duality between the Koopman evolution of observables and the Perron--Frobenius evolution of densities. For volume-preserving flows, this dual structure leads to a unitary representation suitable for quantum simulation. Building on the state-preparation and Liouville-evolution algorithms of Refs.~\cite{Zylberman2024,Zylberman2025,Zylberman2026}, the reconstruction and measurement analysis developed here completes an end-to-end quantum workflow for solving nonlinear ODEs.
We clarified the relationship between the Koopman and Perron--Frobenius viewpoints. While Koopman-based linearization approaches act on a space of observables \cite{Liu2021,novikau2025}, the Perron--Frobenius formulation describes the evolution of probability densities through the Liouville equation. This distinction is central for quantum implementations, it separates linearization in observable space from the probabilistic phase-space embedding.
In the case where the observable space is taken to be $L^2$, the Koopman operator is unitary for volume-preserving flows, and its adjoint provides the corresponding unitary evolution of densities. This dual structure enables a direct comparison between Koopman linearization methods and Liouville-based quantum simulation schemes.
For the Lotka--Volterra system and the quartic oscillator, their Carleman linearized representation converges poorly according to the criterion introduced in Ref.~\cite{Liu2021}. Similar conclusions have been formulated in Ref.\ \cite{novikau2025}. This observation motivates the use of the Liouville formulation, but also highlights the need to understand numerical limitations of both approaches.

 We showed that, once discretization errors of the type discussed in Ref.~\cite{Zylberman2025} are controlled, the dominant source of trajectory-reconstruction error is the folding of the probability density induced by the Hessian of the flow.
This is an extension of previous analyzes of density evolution emphasizing Jacobian-driven deformation effects \cite{JIN2023}.
This result indicates that bounds based only on Jacobian transformations may have limited predictive power for averaged observables, such as the mean position. Indeed, the Jacobian primarily modifies the covariance matrix through a stretching mechanism analogous to state squeezing \cite{Joseph2020}, whereas the Hessian produces nonlinear folding of the density, which directly biases the reconstructed trajectory.
This distinction also clarifies the role of stability in the present framework. Although the transport scheme introduced in Ref.~\cite{Zylberman2025} and reviewed in Sec.~\ref{Sec:Transport} is unconditionally stable at the level of density evolution, an additional stability criterion emerges when the reconstruction of observables from measurement outcomes is considered as detailed in Sec.~\ref{Sec: ODE Lim}.
This observation motivates the measurement-based analysis developed in Sec.~\ref{Sec:Measurement}, where the finite covariance of the transported density is shown to affect both the inferred observable and the associated measurement precision.
These predictions were verified by accurate computations of the quartic oscillator and the Lotka-Volterra systems in Sec.\ \ref{Sec: Numerical results}.
For the quartic oscillator, rapid error growth was observed when the covariance did not meet the required criterion \ref{Eq: Pract crit}.

We proposed a simple measurement strategy for estimating averaged quantities from the Liouville density. This procedure benefit from amplitude estimation techniques, as discussed in Sec.\ \ref{Sec:Measurement} and App.~\ref{App.: Bayes}. The analysis shows that the measurement accuracy is limited not only by sampling complexity, but also by the covariance of the transported density and by the smoothness of the observable being estimated.
Taken together, these results offers a guideline: the initial Gaussian should be chosen as narrow as possible while remaining sufficiently resolved by the computational grid. If the covariance is too large, nonlinear folding rapidly biases the reconstructed mean trajectory. Conversely, if the covariance is too small, the Gaussian becomes under-resolved and grid-scale artifacts introduce additional numerical error, resulting in a growth of the covariance.
Thus, the accuracy of the Liouville embedding is controlled by a balance between covariance width and grid resolution. This trade-off provides a key stability criterion for choosing the initial covariance and for estimating the maximal reliable simulation time.
Moreover, the measurement presented in Sec.\ \ref{Sec:Measurement} offers also an evaluation of the variance, thus enabling a direct monitoring of the criterion.

Finally, Eq ~\ref{Eq: Pract crit} provides a criterion which supports the feasibility of dual KvN implementations for conservative non-linear ODE systems.
Moreover, we designed a mathematical framework to support a detailed analysis of quantum computations of non-linear ODE systems.

\section{Acknowledgement}

The authors would like to thank A. Ameri, P. Bonoli, M. Brachet, P. Cappellaro, H. Krovi,
T. Laakonen, M. Pujol, P. Rall, S. Zhuk and S. Yilmaz for useful discussions on the content of this work. 
The authors acknowledge the MIT Office of Research Computing and Data for providing high performance computing resources that have contributed to the research results reported within this paper.
T. Fredon and N. F. Loureiro were funded by U.S. Department of Energy Grant No. DE-SC0020264 and
by the MIT-IBM Watson AI Lab. The
views expressed are those of the authors, and do not reflect the official policy or position of IBM or the IBM Quantum team.
J.Zylberman 
was supported by France 2030 under the French National
Research Agency award number “ANR-22-PNCQ-0002”.
A. K. Ram was supported by U.S. Department of Energy Grant No. DE–FG02-91ER54109.

\bibliography{Bib-Thib.bib}

\begin{thebibliography}{52}%
\makeatletter
\providecommand \@ifxundefined [1]{%
 \@ifx{#1\undefined}
}%
\providecommand \@ifnum [1]{%
 \ifnum #1\expandafter \@firstoftwo
 \else \expandafter \@secondoftwo
 \fi
}%
\providecommand \@ifx [1]{%
 \ifx #1\expandafter \@firstoftwo
 \else \expandafter \@secondoftwo
 \fi
}%
\providecommand \natexlab [1]{#1}%
\providecommand \enquote  [1]{``#1''}%
\providecommand \bibnamefont  [1]{#1}%
\providecommand \bibfnamefont [1]{#1}%
\providecommand \citenamefont [1]{#1}%
\providecommand \href@noop [0]{\@secondoftwo}%
\providecommand \href [0]{\begingroup \@sanitize@url \@href}%
\providecommand \@href[1]{\@@startlink{#1}\@@href}%
\providecommand \@@href[1]{\endgroup#1\@@endlink}%
\providecommand \@sanitize@url [0]{\catcode `\\12\catcode `\$12\catcode `\&12\catcode `\#12\catcode `\^12\catcode `\_12\catcode `\%12\relax}%
\providecommand \@@startlink[1]{}%
\providecommand \@@endlink[0]{}%
\providecommand \url  [0]{\begingroup\@sanitize@url \@url }%
\providecommand \@url [1]{\endgroup\@href {#1}{\urlprefix }}%
\providecommand \urlprefix  [0]{URL }%
\providecommand \Eprint [0]{\href }%
\providecommand \doibase [0]{https://doi.org/}%
\providecommand \selectlanguage [0]{\@gobble}%
\providecommand \bibinfo  [0]{\@secondoftwo}%
\providecommand \bibfield  [0]{\@secondoftwo}%
\providecommand \translation [1]{[#1]}%
\providecommand \BibitemOpen [0]{}%
\providecommand \bibitemStop [0]{}%
\providecommand \bibitemNoStop [0]{.\EOS\space}%
\providecommand \EOS [0]{\spacefactor3000\relax}%
\providecommand \BibitemShut  [1]{\csname bibitem#1\endcsname}%
\let\auto@bib@innerbib\@empty
%</preamble>
\bibitem [{\citenamefont {Leyton}\ and\ \citenamefont {Osborne}(2008)}]{Leyton09}%
  \BibitemOpen
  \bibfield  {author} {\bibinfo {author} {\bibfnamefont {S.~K.}\ \bibnamefont {Leyton}}\ and\ \bibinfo {author} {\bibfnamefont {T.~J.}\ \bibnamefont {Osborne}},\ }\href {https://doi.org/10.48550/arXiv.0812.4423} {\bibinfo {title} {A quantum algorithm to solve nonlinear differential equations}} (\bibinfo {year} {2008}),\ \bibinfo {note} {arXiv:0812.4423},\ \Eprint {https://arxiv.org/abs/0812.4423} {arXiv:0812.4423 [quant-ph]} \BibitemShut {NoStop}%
\bibitem [{\citenamefont {Liu}\ \emph {et~al.}(2021)\citenamefont {Liu}, \citenamefont {Kolden}, \citenamefont {Krovi}, \citenamefont {Loureiro}, \citenamefont {Trivisa},\ and\ \citenamefont {Childs}}]{Liu2021}%
  \BibitemOpen
  \bibfield  {author} {\bibinfo {author} {\bibfnamefont {J.-P.}\ \bibnamefont {Liu}}, \bibinfo {author} {\bibfnamefont {H.~O.}\ \bibnamefont {Kolden}}, \bibinfo {author} {\bibfnamefont {H.~K.}\ \bibnamefont {Krovi}}, \bibinfo {author} {\bibfnamefont {N.~F.}\ \bibnamefont {Loureiro}}, \bibinfo {author} {\bibfnamefont {K.}~\bibnamefont {Trivisa}},\ and\ \bibinfo {author} {\bibfnamefont {A.~M.}\ \bibnamefont {Childs}},\ }\bibfield  {title} {\bibinfo {title} {Efficient quantum algorithm for dissipative nonlinear differential equations},\ }\href {http://dx.doi.org/10.1073/pnas.2026805118} {\bibfield  {journal} {\bibinfo  {journal} {P NATL ACAD SCI USA}\ }\textbf {\bibinfo {volume} {118}} (\bibinfo {year} {2021})}\BibitemShut {NoStop}%
\bibitem [{\citenamefont {Childs}\ \emph {et~al.}(2021)\citenamefont {Childs}, \citenamefont {Liu},\ and\ \citenamefont {Ostrander}}]{CLO21}%
  \BibitemOpen
  \bibfield  {author} {\bibinfo {author} {\bibfnamefont {A.~M.}\ \bibnamefont {Childs}}, \bibinfo {author} {\bibfnamefont {J.-P.}\ \bibnamefont {Liu}},\ and\ \bibinfo {author} {\bibfnamefont {A.}~\bibnamefont {Ostrander}},\ }\bibfield  {title} {\bibinfo {title} {High-precision quantum algorithms for partial differential equations},\ }\href {https://doi.org/10.22331/q-2021-11-10-574} {\bibfield  {journal} {\bibinfo  {journal} {Quantum}\ }\textbf {\bibinfo {volume} {5}},\ \bibinfo {pages} {574} (\bibinfo {year} {2021})}\BibitemShut {NoStop}%
\bibitem [{\citenamefont {An}\ \emph {et~al.}(2023)\citenamefont {An}, \citenamefont {Childs},\ and\ \citenamefont {Lin}}]{Childs23}%
  \BibitemOpen
  \bibfield  {author} {\bibinfo {author} {\bibfnamefont {D.}~\bibnamefont {An}}, \bibinfo {author} {\bibfnamefont {A.~M.}\ \bibnamefont {Childs}},\ and\ \bibinfo {author} {\bibfnamefont {L.}~\bibnamefont {Lin}},\ }\href {https://doi.org/10.48550/ARXIV.2312.03916} {\bibinfo {title} {Quantum algorithm for linear non-unitary dynamics with near-optimal dependence on all parameters}} (\bibinfo {year} {2023})\BibitemShut {NoStop}%
\bibitem [{\citenamefont {Koukoutsis}\ \emph {et~al.}(2026)\citenamefont {Koukoutsis}, \citenamefont {Hizanidis}, \citenamefont {Gamiz}, \citenamefont {Amaro}, \citenamefont {Tsironis}, \citenamefont {Ram},\ and\ \citenamefont {Vahala}}]{Koukoutsis26}%
  \BibitemOpen
  \bibfield  {author} {\bibinfo {author} {\bibfnamefont {E.}~\bibnamefont {Koukoutsis}}, \bibinfo {author} {\bibfnamefont {K.}~\bibnamefont {Hizanidis}}, \bibinfo {author} {\bibfnamefont {L.~I.~I.}\ \bibnamefont {Gamiz}}, \bibinfo {author} {\bibfnamefont {O.}~\bibnamefont {Amaro}}, \bibinfo {author} {\bibfnamefont {C.}~\bibnamefont {Tsironis}}, \bibinfo {author} {\bibfnamefont {A.~K.}\ \bibnamefont {Ram}},\ and\ \bibinfo {author} {\bibfnamefont {G.}~\bibnamefont {Vahala}},\ }\href {https://doi.org/10.48550/ARXIV.2604.10794} {\bibinfo {title} {Symplectic perspective to quantum computing for hamiltonian systems}} (\bibinfo {year} {2026})\BibitemShut {NoStop}%
\bibitem [{\citenamefont {Lloyd}(1996)}]{Lloyd1996}%
  \BibitemOpen
  \bibfield  {author} {\bibinfo {author} {\bibfnamefont {S.}~\bibnamefont {Lloyd}},\ }\bibfield  {title} {\bibinfo {title} {Universal quantum simulators},\ }\href {https://doi.org/10.1126/science.273.5278.1073} {\bibfield  {journal} {\bibinfo  {journal} {Science}\ }\textbf {\bibinfo {volume} {273}},\ \bibinfo {pages} {1073–1078} (\bibinfo {year} {1996})}\BibitemShut {NoStop}%
\bibitem [{\citenamefont {Low}\ and\ \citenamefont {Chuang}(2017)}]{Low2017}%
  \BibitemOpen
  \bibfield  {author} {\bibinfo {author} {\bibfnamefont {G.~H.}\ \bibnamefont {Low}}\ and\ \bibinfo {author} {\bibfnamefont {I.~L.}\ \bibnamefont {Chuang}},\ }\bibfield  {title} {\bibinfo {title} {Optimal hamiltonian simulation by quantum signal processing},\ }\bibfield  {journal} {\bibinfo  {journal} {PHYS REV LETT}\ }\textbf {\bibinfo {volume} {118}},\ \href {https://doi.org/10.1103/physrevlett.118.010501} {10.1103/physrevlett.118.010501} (\bibinfo {year} {2017})\BibitemShut {NoStop}%
\bibitem [{\citenamefont {Low}\ and\ \citenamefont {Chuang}(2019)}]{Low2019}%
  \BibitemOpen
  \bibfield  {author} {\bibinfo {author} {\bibfnamefont {G.~H.}\ \bibnamefont {Low}}\ and\ \bibinfo {author} {\bibfnamefont {I.~L.}\ \bibnamefont {Chuang}},\ }\bibfield  {title} {\bibinfo {title} {Hamiltonian simulation by qubitization},\ }\href {http://dx.doi.org/10.22331/q-2019-07-12-163} {\bibfield  {journal} {\bibinfo  {journal} {Quantum}\ }\textbf {\bibinfo {volume} {3}},\ \bibinfo {pages} {163} (\bibinfo {year} {2019})}\BibitemShut {NoStop}%
\bibitem [{\citenamefont {Bastidas}\ \emph {et~al.}(2024)\citenamefont {Bastidas}, \citenamefont {Zeytinoğlu}, \citenamefont {Rossi}, \citenamefont {Chuang},\ and\ \citenamefont {Munro}}]{Bastidas2024}%
  \BibitemOpen
  \bibfield  {author} {\bibinfo {author} {\bibfnamefont {V.~M.}\ \bibnamefont {Bastidas}}, \bibinfo {author} {\bibfnamefont {S.}~\bibnamefont {Zeytinoğlu}}, \bibinfo {author} {\bibfnamefont {Z.~M.}\ \bibnamefont {Rossi}}, \bibinfo {author} {\bibfnamefont {I.~L.}\ \bibnamefont {Chuang}},\ and\ \bibinfo {author} {\bibfnamefont {W.~J.}\ \bibnamefont {Munro}},\ }\bibfield  {title} {\bibinfo {title} {Quantum signal processing with the one-dimensional quantum ising model},\ }\href {http://dx.doi.org/10.1103/PhysRevB.109.014306} {\bibfield  {journal} {\bibinfo  {journal} {Physical Review B}\ }\textbf {\bibinfo {volume} {109}} (\bibinfo {year} {2024})}\BibitemShut {NoStop}%
\bibitem [{\citenamefont {Harrow}\ \emph {et~al.}(2009)\citenamefont {Harrow}, \citenamefont {Hassidim},\ and\ \citenamefont {Lloyd}}]{HHL09}%
  \BibitemOpen
  \bibfield  {author} {\bibinfo {author} {\bibfnamefont {A.~W.}\ \bibnamefont {Harrow}}, \bibinfo {author} {\bibfnamefont {A.}~\bibnamefont {Hassidim}},\ and\ \bibinfo {author} {\bibfnamefont {S.}~\bibnamefont {Lloyd}},\ }\bibfield  {title} {\bibinfo {title} {Quantum algorithm for linear systems of equations},\ }\href {https://doi.org/10.1103/physrevlett.103.150502} {\bibfield  {journal} {\bibinfo  {journal} {Phys. Rev. Lett.}\ }\textbf {\bibinfo {volume} {103}} (\bibinfo {year} {2009})}\BibitemShut {NoStop}%
\bibitem [{\citenamefont {Morales}\ \emph {et~al.}(2024)\citenamefont {Morales}, \citenamefont {Pira}, \citenamefont {Schleich}, \citenamefont {Koor}, \citenamefont {Costa}, \citenamefont {An}, \citenamefont {Aspuru-Guzik}, \citenamefont {Lin}, \citenamefont {Rebentrost},\ and\ \citenamefont {Berry}}]{Morales24}%
  \BibitemOpen
  \bibfield  {author} {\bibinfo {author} {\bibfnamefont {M.~E.~S.}\ \bibnamefont {Morales}}, \bibinfo {author} {\bibfnamefont {L.}~\bibnamefont {Pira}}, \bibinfo {author} {\bibfnamefont {P.}~\bibnamefont {Schleich}}, \bibinfo {author} {\bibfnamefont {K.}~\bibnamefont {Koor}}, \bibinfo {author} {\bibfnamefont {P.~C.~S.}\ \bibnamefont {Costa}}, \bibinfo {author} {\bibfnamefont {D.}~\bibnamefont {An}}, \bibinfo {author} {\bibfnamefont {A.}~\bibnamefont {Aspuru-Guzik}}, \bibinfo {author} {\bibfnamefont {L.}~\bibnamefont {Lin}}, \bibinfo {author} {\bibfnamefont {P.}~\bibnamefont {Rebentrost}},\ and\ \bibinfo {author} {\bibfnamefont {D.~W.}\ \bibnamefont {Berry}},\ }\href {https://doi.org/10.48550/ARXIV.2411.02522} {\bibinfo {title} {Quantum linear system solvers: A survey of algorithms and applications}} (\bibinfo {year} {2024})\BibitemShut {NoStop}%
\bibitem [{\citenamefont {Zylberman}\ \emph {et~al.}(2022)\citenamefont {Zylberman}, \citenamefont {{Di Molfetta}}, \citenamefont {Brachet}, \citenamefont {Loureiro},\ and\ \citenamefont {Debbasch}}]{ZDMD22}%
  \BibitemOpen
  \bibfield  {author} {\bibinfo {author} {\bibfnamefont {J.}~\bibnamefont {Zylberman}}, \bibinfo {author} {\bibfnamefont {G.}~\bibnamefont {{Di Molfetta}}}, \bibinfo {author} {\bibfnamefont {M.}~\bibnamefont {Brachet}}, \bibinfo {author} {\bibfnamefont {N.~F.}\ \bibnamefont {Loureiro}},\ and\ \bibinfo {author} {\bibfnamefont {F.}~\bibnamefont {Debbasch}},\ }\bibfield  {title} {\bibinfo {title} {Quantum simulations of hydrodynamics via the madelung transformation},\ }\href {https://doi.org/10.1103/physreva.106.032408} {\bibfield  {journal} {\bibinfo  {journal} {Phys. Rev. A}\ }\textbf {\bibinfo {volume} {106}},\ \bibinfo {pages} {032408} (\bibinfo {year} {2022})}\BibitemShut {NoStop}%
\bibitem [{\citenamefont {Carleman}(1932)}]{Carleman1932}%
  \BibitemOpen
  \bibfield  {author} {\bibinfo {author} {\bibfnamefont {T.}~\bibnamefont {Carleman}},\ }\bibfield  {title} {\bibinfo {title} {Application de la théorie des équations intégrales linéaires aux systèmes d’équations différentielles non linéaires},\ }\href {https://projecteuclid.org/journals/acta-mathematica/volume-59/issue-none/Application-de-la-théorie-des-équations-intégrales-linéaires-aux-systèmes/10.1007/BF02546499.full?tab=ArticleLinkReference} {\bibfield  {journal} {\bibinfo  {journal} {Acta Mathematica}\ }\textbf {\bibinfo {volume} {59}},\ \bibinfo {pages} {63–87} (\bibinfo {year} {1932})}\BibitemShut {NoStop}%
\bibitem [{\citenamefont {May}\ and\ \citenamefont {Qin}(2023)}]{May2023}%
  \BibitemOpen
  \bibfield  {author} {\bibinfo {author} {\bibfnamefont {M.~Q.}\ \bibnamefont {May}}\ and\ \bibinfo {author} {\bibfnamefont {H.}~\bibnamefont {Qin}},\ }\bibfield  {title} {\bibinfo {title} {Quantum three-wave instability},\ }\bibfield  {journal} {\bibinfo  {journal} {Physical Review A}\ }\textbf {\bibinfo {volume} {107}},\ \href {https://doi.org/10.1103/physreva.107.062204} {10.1103/physreva.107.062204} (\bibinfo {year} {2023})\BibitemShut {NoStop}%
\bibitem [{\citenamefont {May}\ and\ \citenamefont {Qin}(2024)}]{may2024}%
  \BibitemOpen
  \bibfield  {author} {\bibinfo {author} {\bibfnamefont {M.~Q.}\ \bibnamefont {May}}\ and\ \bibinfo {author} {\bibfnamefont {H.}~\bibnamefont {Qin}},\ }\bibfield  {title} {\bibinfo {title} {Nonlinear solution of classical three-wave interaction via finite-dimensional quantum model},\ }\bibfield  {journal} {\bibinfo  {journal} {Journal of Plasma Physics}\ }\textbf {\bibinfo {volume} {90}},\ \href {https://doi.org/10.1017/s002237782400059x} {10.1017/s002237782400059x} (\bibinfo {year} {2024})\BibitemShut {NoStop}%
\bibitem [{\citenamefont {May}(2025)}]{MayPHD}%
  \BibitemOpen
  \bibfield  {author} {\bibinfo {author} {\bibfnamefont {M.}~\bibnamefont {May}},\ }\emph {\bibinfo {title} {Quantum Computing for Plasma Physics via Second Quantization}},\ \href {https://www.proquest.com/dissertations-theses/quantum-computing-plasma-physics-via-second/docview/3256676255/se-2} {Ph.D. thesis} (\bibinfo {year} {2025})\BibitemShut {NoStop}%
\bibitem [{\citenamefont {Andress}\ \emph {et~al.}(2024)\citenamefont {Andress}, \citenamefont {Engel}, \citenamefont {Shi},\ and\ \citenamefont {Parker}}]{Andress2024}%
  \BibitemOpen
  \bibfield  {author} {\bibinfo {author} {\bibfnamefont {J.}~\bibnamefont {Andress}}, \bibinfo {author} {\bibfnamefont {A.}~\bibnamefont {Engel}}, \bibinfo {author} {\bibfnamefont {Y.}~\bibnamefont {Shi}},\ and\ \bibinfo {author} {\bibfnamefont {S.}~\bibnamefont {Parker}},\ }\href {https://doi.org/10.48550/ARXIV.2410.03838} {\bibinfo {title} {Quantum simulation of nonlinear dynamical systems using repeated measurement}} (\bibinfo {year} {2024})\BibitemShut {NoStop}%
\bibitem [{\citenamefont {Shi}\ \emph {et~al.}(2024)\citenamefont {Shi}, \citenamefont {Evert}, \citenamefont {Brown}, \citenamefont {Tripathi}, \citenamefont {Sete}, \citenamefont {Geyko}, \citenamefont {Cho}, \citenamefont {DuBois}, \citenamefont {Lidar}, \citenamefont {Joseph},\ and\ \citenamefont {Reagor}}]{Shi2024}%
  \BibitemOpen
  \bibfield  {author} {\bibinfo {author} {\bibfnamefont {Y.}~\bibnamefont {Shi}}, \bibinfo {author} {\bibfnamefont {B.}~\bibnamefont {Evert}}, \bibinfo {author} {\bibfnamefont {A.~F.}\ \bibnamefont {Brown}}, \bibinfo {author} {\bibfnamefont {V.}~\bibnamefont {Tripathi}}, \bibinfo {author} {\bibfnamefont {E.~A.}\ \bibnamefont {Sete}}, \bibinfo {author} {\bibfnamefont {V.}~\bibnamefont {Geyko}}, \bibinfo {author} {\bibfnamefont {Y.}~\bibnamefont {Cho}}, \bibinfo {author} {\bibfnamefont {J.~L.}\ \bibnamefont {DuBois}}, \bibinfo {author} {\bibfnamefont {D.}~\bibnamefont {Lidar}}, \bibinfo {author} {\bibfnamefont {I.}~\bibnamefont {Joseph}},\ and\ \bibinfo {author} {\bibfnamefont {M.}~\bibnamefont {Reagor}},\ }\bibfield  {title} {\bibinfo {title} {Simulating nonlinear optical processes on a superconducting quantum device},\ }\bibfield  {journal} {\bibinfo  {journal} {Journal of Plasma Physics}\ }\textbf {\bibinfo {volume} {90}},\ \href {https://doi.org/10.1017/s0022377824001326} {10.1017/s0022377824001326}
  (\bibinfo {year} {2024})\BibitemShut {NoStop}%
\bibitem [{\citenamefont {Novikau}\ and\ \citenamefont {Joseph}(2024)}]{novikau2024}%
  \BibitemOpen
  \bibfield  {author} {\bibinfo {author} {\bibfnamefont {I.}~\bibnamefont {Novikau}}\ and\ \bibinfo {author} {\bibfnamefont {I.}~\bibnamefont {Joseph}},\ }\href {https://arxiv.org/abs/2410.03985} {\bibinfo {title} {Quantum algorithm for the advection-diffusion equation and the koopman-von neumann approach to nonlinear dynamical systems}} (\bibinfo {year} {2024}),\ \Eprint {https://arxiv.org/abs/2410.03985} {arXiv:2410.03985 [physics.comp-ph]} \BibitemShut {NoStop}%
\bibitem [{\citenamefont {Zylberman}\ \emph {et~al.}(2026{\natexlab{a}})\citenamefont {Zylberman}, \citenamefont {Fredon}, \citenamefont {Loureiro},\ and\ \citenamefont {Debbasch}}]{Zylberman2026}%
  \BibitemOpen
  \bibfield  {author} {\bibinfo {author} {\bibfnamefont {J.}~\bibnamefont {Zylberman}}, \bibinfo {author} {\bibfnamefont {T.}~\bibnamefont {Fredon}}, \bibinfo {author} {\bibfnamefont {N.~F.}\ \bibnamefont {Loureiro}},\ and\ \bibinfo {author} {\bibfnamefont {F.}~\bibnamefont {Debbasch}},\ }\bibinfo {title} {Quantum algorithm for anisotropic diffusion and convection equations with vector norm scaling},\ in\ \href {https://doi.org/10.1007/978-3-032-13855-2_23} {\emph {\bibinfo {booktitle} {Quantum Engineering Sciences and Technologies for Industry and Services}}}\ (\bibinfo  {publisher} {Springer Nature Switzerland},\ \bibinfo {year} {2026})\ p.\ \bibinfo {pages} {255–263}\BibitemShut {NoStop}%
\bibitem [{\citenamefont {Joseph}\ \emph {et~al.}(2023)\citenamefont {Joseph}, \citenamefont {Shi}, \citenamefont {Porter}, \citenamefont {Castelli}, \citenamefont {Geyko}, \citenamefont {Graziani}, \citenamefont {Libby},\ and\ \citenamefont {DuBois}}]{Joseph2023}%
  \BibitemOpen
  \bibfield  {author} {\bibinfo {author} {\bibfnamefont {I.}~\bibnamefont {Joseph}}, \bibinfo {author} {\bibfnamefont {Y.}~\bibnamefont {Shi}}, \bibinfo {author} {\bibfnamefont {M.~D.}\ \bibnamefont {Porter}}, \bibinfo {author} {\bibfnamefont {A.~R.}\ \bibnamefont {Castelli}}, \bibinfo {author} {\bibfnamefont {V.~I.}\ \bibnamefont {Geyko}}, \bibinfo {author} {\bibfnamefont {F.~R.}\ \bibnamefont {Graziani}}, \bibinfo {author} {\bibfnamefont {S.~B.}\ \bibnamefont {Libby}},\ and\ \bibinfo {author} {\bibfnamefont {J.~L.}\ \bibnamefont {DuBois}},\ }\bibfield  {title} {\bibinfo {title} {Quantum computing for fusion energy science applications},\ }\bibfield  {journal} {\bibinfo  {journal} {Physics of Plasmas}\ }\textbf {\bibinfo {volume} {30}},\ \href {https://doi.org/10.1063/5.0123765} {10.1063/5.0123765} (\bibinfo {year} {2023})\BibitemShut {NoStop}%
\bibitem [{\citenamefont {Koopman}(1931)}]{Koopman31}%
  \BibitemOpen
  \bibfield  {author} {\bibinfo {author} {\bibfnamefont {B.~O.}\ \bibnamefont {Koopman}},\ }\bibfield  {title} {\bibinfo {title} {Hamiltonian systems and transformation in hilbert space},\ }\href {https://doi.org/10.1073/pnas.17.5.315} {\bibfield  {journal} {\bibinfo  {journal} {P NATL ACAD SCI USA}\ }\textbf {\bibinfo {volume} {17}},\ \bibinfo {pages} {315} (\bibinfo {year} {1931})}\BibitemShut {NoStop}%
\bibitem [{\citenamefont {Klein}(2017)}]{Klein2017}%
  \BibitemOpen
  \bibfield  {author} {\bibinfo {author} {\bibfnamefont {U.}~\bibnamefont {Klein}},\ }\bibfield  {title} {\bibinfo {title} {From koopman–von neumann theory to quantum theory},\ }\href {http://dx.doi.org/10.1007/s40509-017-0113-2} {\bibfield  {journal} {\bibinfo  {journal} {Quantum Stud.: Math. Found.}\ }\textbf {\bibinfo {volume} {5}},\ \bibinfo {pages} {219–227} (\bibinfo {year} {2017})}\BibitemShut {NoStop}%
\bibitem [{\citenamefont {Mezi{\'c}}(2005)}]{Mezic2005}%
  \BibitemOpen
  \bibfield  {author} {\bibinfo {author} {\bibfnamefont {I.}~\bibnamefont {Mezi{\'c}}},\ }\bibfield  {title} {\bibinfo {title} {Spectral properties of dynamical systems, model reduction and decompositions},\ }\href {https://doi.org/10.1007/s11071-005-2824-x} {\bibfield  {journal} {\bibinfo  {journal} {Nonlinear Dynamics}\ }\textbf {\bibinfo {volume} {41}},\ \bibinfo {pages} {309} (\bibinfo {year} {2005})}\BibitemShut {NoStop}%
\bibitem [{\citenamefont {Budi{\v{s}}i{\'c}}\ \emph {et~al.}(2012)\citenamefont {Budi{\v{s}}i{\'c}}, \citenamefont {Mohr},\ and\ \citenamefont {Mezi{\'c}}}]{Budisic2012}%
  \BibitemOpen
  \bibfield  {author} {\bibinfo {author} {\bibfnamefont {M.}~\bibnamefont {Budi{\v{s}}i{\'c}}}, \bibinfo {author} {\bibfnamefont {R.}~\bibnamefont {Mohr}},\ and\ \bibinfo {author} {\bibfnamefont {I.}~\bibnamefont {Mezi{\'c}}},\ }\bibfield  {title} {\bibinfo {title} {Applied koopmanism},\ }\href {https://doi.org/10.1063/1.4772195} {\bibfield  {journal} {\bibinfo  {journal} {Chaos}\ }\textbf {\bibinfo {volume} {22}},\ \bibinfo {pages} {047510} (\bibinfo {year} {2012})}\BibitemShut {NoStop}%
\bibitem [{\citenamefont {Mezi{\'c}}(2013)}]{Mezic2013}%
  \BibitemOpen
  \bibfield  {author} {\bibinfo {author} {\bibfnamefont {I.}~\bibnamefont {Mezi{\'c}}},\ }\bibfield  {title} {\bibinfo {title} {Analysis of fluid flows via spectral properties of the koopman operator},\ }\href {https://doi.org/10.1146/annurev-fluid-011212-140652} {\bibfield  {journal} {\bibinfo  {journal} {Annual Review of Fluid Mechanics}\ }\textbf {\bibinfo {volume} {45}},\ \bibinfo {pages} {357} (\bibinfo {year} {2013})}\BibitemShut {NoStop}%
\bibitem [{\citenamefont {Klus}\ \emph {et~al.}(2016)\citenamefont {Klus}, \citenamefont {Koltai},\ and\ \citenamefont {Sch\"{u}tte}}]{Klus2016}%
  \BibitemOpen
  \bibfield  {author} {\bibinfo {author} {\bibfnamefont {S.}~\bibnamefont {Klus}}, \bibinfo {author} {\bibfnamefont {P.}~\bibnamefont {Koltai}},\ and\ \bibinfo {author} {\bibfnamefont {C.}~\bibnamefont {Sch\"{u}tte}},\ }\bibfield  {title} {\bibinfo {title} {On the numerical approximation of the perron-frobenius and koopman operator},\ }\href {https://doi.org/10.3934/jcd.2016003} {\bibfield  {journal} {\bibinfo  {journal} {Journal of Computational Dynamics}\ }\textbf {\bibinfo {volume} {3}},\ \bibinfo {pages} {51–79} (\bibinfo {year} {2016})}\BibitemShut {NoStop}%
\bibitem [{\citenamefont {Mauroy}\ \emph {et~al.}(2020)\citenamefont {Mauroy}, \citenamefont {Mezic},\ and\ \citenamefont {Susuki}}]{MMS20}%
  \BibitemOpen
  \bibfield  {author} {\bibinfo {author} {\bibnamefont {Mauroy}}, \bibinfo {author} {\bibnamefont {Mezic}},\ and\ \bibinfo {author} {\bibnamefont {Susuki}},\ }\href {https://doi.org/10.1007/978-3-030-35713-9} {\emph {\bibinfo {title} {The Koopman Operator in Systems and Control: Concepts, Methodologies, and Applications}}}\ (\bibinfo  {publisher} {Springer International Publishing},\ \bibinfo {year} {2020})\BibitemShut {NoStop}%
\bibitem [{\citenamefont {Joseph}(2020)}]{Joseph2020}%
  \BibitemOpen
  \bibfield  {author} {\bibinfo {author} {\bibfnamefont {I.}~\bibnamefont {Joseph}},\ }\bibfield  {title} {\bibinfo {title} {Koopman–von neumann approach to quantum simulation of nonlinear classical dynamics},\ }\href {http://dx.doi.org/10.1103/PhysRevResearch.2.043102} {\bibfield  {journal} {\bibinfo  {journal} {PHYS RES}\ }\textbf {\bibinfo {volume} {2}} (\bibinfo {year} {2020})}\BibitemShut {NoStop}%
\bibitem [{\citenamefont {Zylberman}\ \emph {et~al.}(2026{\natexlab{b}})\citenamefont {Zylberman}, \citenamefont {Fredon}, \citenamefont {Loureiro},\ and\ \citenamefont {Debbasch}}]{Zylberman2025}%
  \BibitemOpen
  \bibfield  {author} {\bibinfo {author} {\bibfnamefont {J.}~\bibnamefont {Zylberman}}, \bibinfo {author} {\bibfnamefont {T.}~\bibnamefont {Fredon}}, \bibinfo {author} {\bibfnamefont {N.~F.}\ \bibnamefont {Loureiro}},\ and\ \bibinfo {author} {\bibfnamefont {F.}~\bibnamefont {Debbasch}},\ }\bibfield  {title} {\bibinfo {title} {Trotter-based quantum algorithm for solving transport equations with exponentially fewer time-steps},\ }\href {https://doi.org/10.1088/2058-9565/ae488e} {\bibfield  {journal} {\bibinfo  {journal} {Quantum Science and Technology}\ }\textbf {\bibinfo {volume} {11}},\ \bibinfo {pages} {025015} (\bibinfo {year} {2026}{\natexlab{b}})}\BibitemShut {NoStop}%
\bibitem [{\citenamefont {Tanaka}\ and\ \citenamefont {Fujii}(2025)}]{tanaka2025}%
  \BibitemOpen
  \bibfield  {author} {\bibinfo {author} {\bibfnamefont {Y.}~\bibnamefont {Tanaka}}\ and\ \bibinfo {author} {\bibfnamefont {K.}~\bibnamefont {Fujii}},\ }\href {https://arxiv.org/abs/2305.00653} {\bibinfo {title} {A polynomial time quantum algorithm for exponentially large scale nonlinear differential equations via hamiltonian simulation}} (\bibinfo {year} {2025}),\ \Eprint {https://arxiv.org/abs/2305.00653} {arXiv:2305.00653 [quant-ph]} \BibitemShut {NoStop}%
\bibitem [{\citenamefont {Novikau}\ and\ \citenamefont {Joseph}(2025)}]{novikau2025}%
  \BibitemOpen
  \bibfield  {author} {\bibinfo {author} {\bibfnamefont {I.}~\bibnamefont {Novikau}}\ and\ \bibinfo {author} {\bibfnamefont {I.}~\bibnamefont {Joseph}},\ }\href {https://arxiv.org/abs/2510.15715} {\bibinfo {title} {Globalizing the carleman linear embedding method for nonlinear dynamics}} (\bibinfo {year} {2025}),\ \Eprint {https://arxiv.org/abs/2510.15715} {arXiv:2510.15715 [quant-ph]} \BibitemShut {NoStop}%
\bibitem [{\citenamefont {Riesz}(1909)}]{Riesz1909}%
  \BibitemOpen
  \bibfield  {author} {\bibinfo {author} {\bibfnamefont {F.}~\bibnamefont {Riesz}},\ }\bibfield  {title} {\bibinfo {title} {Sur les opérations fonctionnelles linéaires},\ }\href@noop {} {\bibfield  {journal} {\bibinfo  {journal} {Comptes rendus hebdomadaires des séances de l'Académie des sciences}\ }\textbf {\bibinfo {volume} {149}},\ \bibinfo {pages} {974} (\bibinfo {year} {1909})}\BibitemShut {NoStop}%
\bibitem [{\citenamefont {Radon}(1913)}]{Radon1913}%
  \BibitemOpen
  \bibfield  {author} {\bibinfo {author} {\bibfnamefont {J.}~\bibnamefont {Radon}},\ }\bibfield  {title} {\bibinfo {title} {Theorie und anwendungen der absolut additiven mengenfunktionen},\ }\href@noop {} {\bibfield  {journal} {\bibinfo  {journal} {Sitzungsberichte der Kaiserlichen Akademie der Wissenschaften in Wien, Mathematisch-Naturwissenschaftliche Klasse, Abteilung IIa}\ }\textbf {\bibinfo {volume} {122}},\ \bibinfo {pages} {1295} (\bibinfo {year} {1913})}\BibitemShut {NoStop}%
\bibitem [{\citenamefont {Nikodym}(1930)}]{Nikodym1930}%
  \BibitemOpen
  \bibfield  {author} {\bibinfo {author} {\bibfnamefont {O.}~\bibnamefont {Nikodym}},\ }\bibfield  {title} {\bibinfo {title} {Sur une généralisation des intégrales de m. j. radon},\ }\href {https://doi.org/10.4064/fm-15-1-131-179} {\bibfield  {journal} {\bibinfo  {journal} {Fundamenta Mathematicae}\ }\textbf {\bibinfo {volume} {15}},\ \bibinfo {pages} {131} (\bibinfo {year} {1930})}\BibitemShut {NoStop}%
\bibitem [{\citenamefont {LaSalle}(1972)}]{LaSalle1972}%
  \BibitemOpen
  \bibfield  {author} {\bibinfo {author} {\bibfnamefont {J.}~\bibnamefont {LaSalle}},\ }\bibinfo {title} {Dissipative systems},\ in\ \href {https://doi.org/10.1016/b978-0-12-743650-0.50019-8} {\emph {\bibinfo {booktitle} {Ordinary Differential Equations}}}\ (\bibinfo  {publisher} {Elsevier},\ \bibinfo {year} {1972})\ p.\ \bibinfo {pages} {165–174}\BibitemShut {NoStop}%
\bibitem [{\citenamefont {Liu}\ and\ \citenamefont {Jin}(2023)}]{Liu2023}%
  \BibitemOpen
  \bibfield  {author} {\bibinfo {author} {\bibfnamefont {N.}~\bibnamefont {Liu}}\ and\ \bibinfo {author} {\bibfnamefont {S.}~\bibnamefont {Jin}},\ }\bibfield  {title} {\bibinfo {title} {Efficient quantum algorithms for dissipative systems},\ }\href {https://doi.org/10.1103/PhysRevLett.130.160602} {\bibfield  {journal} {\bibinfo  {journal} {Physical Review Letters}\ }\textbf {\bibinfo {volume} {130}},\ \bibinfo {pages} {160602} (\bibinfo {year} {2023})},\ \Eprint {https://arxiv.org/abs/2212.13923} {arXiv:2212.13923 [quant-ph]} \BibitemShut {NoStop}%
\bibitem [{\citenamefont {Zylberman}\ \emph {et~al.}(2024)\citenamefont {Zylberman}, \citenamefont {Nzongani}, \citenamefont {Simonetto},\ and\ \citenamefont {Debbasch}}]{ZNSD24}%
  \BibitemOpen
  \bibfield  {author} {\bibinfo {author} {\bibfnamefont {J.}~\bibnamefont {Zylberman}}, \bibinfo {author} {\bibfnamefont {U.}~\bibnamefont {Nzongani}}, \bibinfo {author} {\bibfnamefont {A.}~\bibnamefont {Simonetto}},\ and\ \bibinfo {author} {\bibfnamefont {F.}~\bibnamefont {Debbasch}},\ }\href {https://doi.org/10.48550/ARXIV.2404.02819} {\bibinfo {title} {Efficient quantum circuits for non-unitary and unitary diagonal operators with space-time-accuracy trade-offs}} (\bibinfo {year} {2024})\BibitemShut {NoStop}%
\bibitem [{\citenamefont {Trotter}(1959)}]{Trotter1959}%
  \BibitemOpen
  \bibfield  {author} {\bibinfo {author} {\bibfnamefont {H.~F.}\ \bibnamefont {Trotter}},\ }\bibfield  {title} {\bibinfo {title} {On the product of semi-groups of operators},\ }\href {http://dx.doi.org/10.1090/S0002-9939-1959-0108732-6} {\bibfield  {journal} {\bibinfo  {journal} {P AM MATH SOC}\ }\textbf {\bibinfo {volume} {10}},\ \bibinfo {pages} {545–551} (\bibinfo {year} {1959})}\BibitemShut {NoStop}%
\bibitem [{\citenamefont {SUZUKI}(1993)}]{Suzuki1993}%
  \BibitemOpen
  \bibfield  {author} {\bibinfo {author} {\bibfnamefont {M.}~\bibnamefont {SUZUKI}},\ }\bibfield  {title} {\bibinfo {title} {General decomposition theory of ordered exponentials.},\ }\href {http://dx.doi.org/10.2183/pjab.69.161} {\bibfield  {journal} {\bibinfo  {journal} {P JPN ACAD B}\ }\textbf {\bibinfo {volume} {69}},\ \bibinfo {pages} {161–166} (\bibinfo {year} {1993})}\BibitemShut {NoStop}%
\bibitem [{\citenamefont {Brassard}\ \emph {et~al.}(2002)\citenamefont {Brassard}, \citenamefont {H{\o}yer}, \citenamefont {Mosca},\ and\ \citenamefont {Tapp}}]{Brassard2002}%
  \BibitemOpen
  \bibfield  {author} {\bibinfo {author} {\bibfnamefont {G.}~\bibnamefont {Brassard}}, \bibinfo {author} {\bibfnamefont {P.}~\bibnamefont {H{\o}yer}}, \bibinfo {author} {\bibfnamefont {M.}~\bibnamefont {Mosca}},\ and\ \bibinfo {author} {\bibfnamefont {A.}~\bibnamefont {Tapp}},\ }\href {https://doi.org/10.1090/conm/305/05215} {\bibinfo {title} {Quantum amplitude amplification and estimation}} (\bibinfo {year} {2002})\BibitemShut {NoStop}%
\bibitem [{\citenamefont {Rall}(2020)}]{Rall2020}%
  \BibitemOpen
  \bibfield  {author} {\bibinfo {author} {\bibfnamefont {P.}~\bibnamefont {Rall}},\ }\bibfield  {title} {\bibinfo {title} {Quantum algorithms for estimating physical quantities using block encodings},\ }\bibfield  {journal} {\bibinfo  {journal} {Physical Review A}\ }\textbf {\bibinfo {volume} {102}},\ \href {https://doi.org/10.1103/physreva.102.022408} {10.1103/physreva.102.022408} (\bibinfo {year} {2020})\BibitemShut {NoStop}%
\bibitem [{\citenamefont {Rall}(2021)}]{Rall2021}%
  \BibitemOpen
  \bibfield  {author} {\bibinfo {author} {\bibfnamefont {P.}~\bibnamefont {Rall}},\ }\bibfield  {title} {\bibinfo {title} {Faster coherent quantum algorithms for phase, energy, and amplitude estimation},\ }\href {https://doi.org/10.22331/q-2021-10-19-566} {\bibfield  {journal} {\bibinfo  {journal} {Quantum}\ }\textbf {\bibinfo {volume} {5}},\ \bibinfo {pages} {566} (\bibinfo {year} {2021})}\BibitemShut {NoStop}%
\bibitem [{\citenamefont {Grover}(1996)}]{Grover96}%
  \BibitemOpen
  \bibfield  {author} {\bibinfo {author} {\bibfnamefont {L.~K.}\ \bibnamefont {Grover}},\ }\bibfield  {title} {\bibinfo {title} {A fast quantum mechanical algorithm for database search},\ }in\ \href {http://dl.acm.org/citation.cfm?doid=237814.237866} {\emph {\bibinfo {booktitle} {Proc. 28th Ann. ACM Symp. on Theory of Computing - STOC96}}}\ (\bibinfo {year} {1996})\BibitemShut {NoStop}%
\bibitem [{\citenamefont {Zylberman}\ and\ \citenamefont {Debbasch}(2024)}]{Zylberman2024}%
  \BibitemOpen
  \bibfield  {author} {\bibinfo {author} {\bibfnamefont {J.}~\bibnamefont {Zylberman}}\ and\ \bibinfo {author} {\bibfnamefont {F.}~\bibnamefont {Debbasch}},\ }\bibfield  {title} {\bibinfo {title} {Efficient quantum state preparation with walsh series},\ }\href {http://dx.doi.org/10.1103/PhysRevA.109.042401} {\bibfield  {journal} {\bibinfo  {journal} {Phys Rev A}\ }\textbf {\bibinfo {volume} {109}} (\bibinfo {year} {2024})}\BibitemShut {NoStop}%
\bibitem [{\citenamefont {Jin}\ \emph {et~al.}(2023)\citenamefont {Jin}, \citenamefont {Liu},\ and\ \citenamefont {Yu}}]{JIN2023}%
  \BibitemOpen
  \bibfield  {author} {\bibinfo {author} {\bibfnamefont {S.}~\bibnamefont {Jin}}, \bibinfo {author} {\bibfnamefont {N.}~\bibnamefont {Liu}},\ and\ \bibinfo {author} {\bibfnamefont {Y.}~\bibnamefont {Yu}},\ }\bibfield  {title} {\bibinfo {title} {Time complexity analysis of quantum algorithms via linear representations for nonlinear ordinary and partial differential equations},\ }\href {https://doi.org/https://doi.org/10.1016/j.jcp.2023.112149} {\bibfield  {journal} {\bibinfo  {journal} {Journal of Computational Physics}\ }\textbf {\bibinfo {volume} {487}},\ \bibinfo {pages} {112149} (\bibinfo {year} {2023})}\BibitemShut {NoStop}%
\bibitem [{\citenamefont {Lichtenberg}\ and\ \citenamefont {Lieberman}(1983)}]{Lichtenberg1983}%
  \BibitemOpen
  \bibfield  {author} {\bibinfo {author} {\bibfnamefont {A.~J.}\ \bibnamefont {Lichtenberg}}\ and\ \bibinfo {author} {\bibfnamefont {M.~A.}\ \bibnamefont {Lieberman}},\ }\href {https://doi.org/10.1007/978-1-4757-4257-2} {\emph {\bibinfo {title} {Regular and Stochastic Motion}}}\ (\bibinfo  {publisher} {Springer New York},\ \bibinfo {year} {1983})\BibitemShut {NoStop}%
\bibitem [{\citenamefont {Nicholson}(1983)}]{Nicholson83}%
  \BibitemOpen
  \bibfield  {author} {\bibinfo {author} {\bibfnamefont {D.~R.}\ \bibnamefont {Nicholson}},\ }\href {https://www.semanticscholar.org/paper/Introduction-to-Plasma-Theory-Nicholson/de7ff83029fdd5b4d1526046a51e0c22d2184636} {\emph {\bibinfo {title} {Introduction to plasma theory}}},\ Vol.~\bibinfo {volume} {1}\ (\bibinfo  {publisher} {Wiley New York},\ \bibinfo {year} {1983})\BibitemShut {NoStop}%
\bibitem [{\citenamefont {David}(2019)}]{david:hal-04438965}%
  \BibitemOpen
  \bibfield  {author} {\bibinfo {author} {\bibfnamefont {F.}~\bibnamefont {David}},\ }\href {https://hal.science/hal-04438965} {\emph {\bibinfo {title} {{Th{\'e}orie Statistique des Champs Tome 1}}}},\ Savoirs Actuels\ (\bibinfo  {publisher} {{EDP Sciences / CNRS Editions}},\ \bibinfo {year} {2019})\BibitemShut {NoStop}%
\bibitem [{\citenamefont {Cocco}\ \emph {et~al.}(2022)\citenamefont {Cocco}, \citenamefont {Monasson},\ and\ \citenamefont {Zamponi}}]{Cocco2022}%
  \BibitemOpen
  \bibfield  {author} {\bibinfo {author} {\bibfnamefont {S.}~\bibnamefont {Cocco}}, \bibinfo {author} {\bibfnamefont {R.}~\bibnamefont {Monasson}},\ and\ \bibinfo {author} {\bibfnamefont {F.}~\bibnamefont {Zamponi}},\ }\href {https://doi.org/10.1093/oso/9780198864745.001.0001} {\emph {\bibinfo {title} {From Statistical Physics to Data-Driven Modelling: with Applications to Quantitative Biology}}}\ (\bibinfo  {publisher} {Oxford University PressOxford},\ \bibinfo {year} {2022})\BibitemShut {NoStop}%
\bibitem [{\citenamefont {Aaronson}\ and\ \citenamefont {Rall}(2020)}]{Aaronson2020}%
  \BibitemOpen
  \bibfield  {author} {\bibinfo {author} {\bibfnamefont {S.}~\bibnamefont {Aaronson}}\ and\ \bibinfo {author} {\bibfnamefont {P.}~\bibnamefont {Rall}},\ }\bibinfo {title} {Quantum approximate counting, simplified},\ in\ \href {https://doi.org/10.1137/1.9781611976014.5} {\emph {\bibinfo {booktitle} {Symposium on Simplicity in Algorithms}}}\ (\bibinfo  {publisher} {Society for Industrial and Applied Mathematics},\ \bibinfo {year} {2020})\ p.\ \bibinfo {pages} {24–32}\BibitemShut {NoStop}%
\bibitem [{\citenamefont {Welch}\ \emph {et~al.}(2014)\citenamefont {Welch}, \citenamefont {Greenbaum}, \citenamefont {Mostame},\ and\ \citenamefont {Aspuru-Guzik}}]{WGM14}%
  \BibitemOpen
  \bibfield  {author} {\bibinfo {author} {\bibfnamefont {J.}~\bibnamefont {Welch}}, \bibinfo {author} {\bibfnamefont {D.}~\bibnamefont {Greenbaum}}, \bibinfo {author} {\bibfnamefont {S.}~\bibnamefont {Mostame}},\ and\ \bibinfo {author} {\bibfnamefont {A.}~\bibnamefont {Aspuru-Guzik}},\ }\bibfield  {title} {\bibinfo {title} {Efficient quantum circuits for diagonal unitaries without ancillas},\ }\href {https://doi.org/10.1088/1367-2630/16/3/033040} {\bibfield  {journal} {\bibinfo  {journal} {New J. Phys.}\ }\textbf {\bibinfo {volume} {16}},\ \bibinfo {pages} {033040} (\bibinfo {year} {2014})}\BibitemShut {NoStop}%
\end{thebibliography}%
\appendix
\section{Hamiltonian formalism}
\label{App.: Ham}
A Hamiltonian system is a dynamical system with a conserved quantity called the Hamiltonian denoted $H$ associated with $N$- coupled variables $((q_i, p_i))_{i\in[\![1,N]\!]}$ \cite{Lichtenberg1983, Nicholson83, david:hal-04438965}.
The equations of motions for the variables $q_i$ and $p_i$ are given by Hamilton equations. 
\begin{subequations}
    \begin{eqnarray}
        d_tq_i = \partial_{p_i}H \\
        d_tp_i =-\partial_{q_i}H 
\end{eqnarray}
\label{Eq: Hamilton eq}
\end{subequations}

where $d_t$ denotes the standard derivative with respect to the variable $t$ and $\partial_x$ is the partial derivative with respect to the variable $x$.

In this description, a state of the system is given by its coordinates$(\boldsymbol q,\boldsymbol p)=(q_i, p_i)_{i\in[\![1,N]\!]}$.
The ensemble of state forms a manifold $\Omega$ that is called the Phase Space \cite{Lichtenberg1983, david:hal-04438965}.

Let $f(\boldsymbol q,\boldsymbol p) $ being an observable \textit{i.e.} a regular scalar function defined over the phase space of the system\footnote{While $f$ can explicitly depend on time, we will assume that not to be the case.}.
The evolution of $f$ is delivered by the chain rule and the Eq.\ \refeq{Eq: Hamilton eq} as
\begin{equation}
    d_t f = \sum \limits_{i=1}^{N} d_tq_i \,\partial_{q_i}f + d_tp_i \,\partial_{p_i}f =  \{f, H\}
    \label{Eq: Observable eq}
\end{equation}

where the Poisson bracket $\{f, H\}$ is defined as $\{f, H\}=\sum \limits_{i=1}^{N} \partial_{p_i} H\,\partial_{q_i}f - \partial_{q_i}H \, \partial_{p_i}f$  \cite{Lichtenberg1983, david:hal-04438965}.

Similarly as in Sec.\ \ref{Sec: ODE Lim}, two trivial examples of such an observable are the characteristics of the movement $q_i:\Omega\rightarrow\mathbb{R}$ and $p_i::\Omega\rightarrow\mathbb{R}$ or the Hamiltonian itself.
For the characteristics, Eq.\ \ref{Eq: Observable eq} boils down to Eq.\ \ref{Eq: Hamilton eq}.
For the Hamiltonian, Eq.\ \ref{Eq: Observable eq} turns out to express the conservation of the Hamiltonian over time.
The latter implies that the evolution of the dynamical system, or its trajectory, is identified by the associated initial value of the Hamiltonian.
This conservation equation is just a manifestation of the Liouville Theorem.

Due to the volume conservation property of the phase space ensured by Liouville theorem, a statistical state $\rho(\textbf{q}, \textbf{p},t)$ can be seen as a mixture of initial conditions can be defined.
Liouville theorem ensures that over time $\rho$ will remain normalized.
The fraction of the state in a part of the Phase Space $\mathcal{P}$ is defined as $\int_{\mathcal{P}}\rho(\textbf{q}, \textbf{p},t) d\textbf{q} d\textbf{p} $ where $d\textbf{q} d\textbf{p} = \prod\limits_{i=1}^{N} dq_i dp_i$ is the phase space volume element.
Similarly as Sec.\ \ref{Sec: KvN}, the existence of the measure is guaranteed by the Riesz Theorem \cite{Riesz1909}.
The measure can be further represented by a density function which is the statistical state if the measure is absolutely continuous according to Radon-Nikodym theorem \cite{Radon1913,Nikodym1930}.

\subsection{Liouville equation}

Either for the Hamiltonian systems or for the class of systems in Sec.\ \ref{Sec: KvN} under the criterion presented in Eq.\ \ref{Eq: Numerical criterion}, the evolution of the number of states in the sub-volume $\mathcal{P}$ over time is given by Stokes theorem

\begin{eqnarray*}
    d_t\int_{\mathcal{P}}\rho(\textbf{q}, \textbf{p}) d\textbf{q} d\textbf{p} &\\=-\sum_{i=1}&\int_{\mathcal{P}} \left(\partial_{q_i}\left(\rho d_tq_i\right)+\partial_{p_i}\left(\rho d_tp_i\right) \right) d\textbf{q} d\textbf{p}\\
    =-\sum_{i=1}&\int_{\mathcal{P}} \left(\partial_{q_i}\left(\rho \partial_{p_i}\right)-\partial_{p_i}\left(\rho \partial_{q_i}H\right) \right) d\textbf{q} d\textbf{p}.
    \nonumber
\end{eqnarray*}

Since $\rho$ is regular and the phase space volume element is conserved, we can put the time derivative under the integral.
Moreover, this derivation is valid for any subpart of the Phase Space and the Hamiltonian verifies the Schwartz theorem (alternatively the differential flow $F_i$ verifies Eq.\ \ref{Eq: Numerical criterion}).
As a result, we can conclude that $\rho$ must verify the Liouville equation

\begin{equation}
    \partial_t \rho = \{H,\, \rho\}= \mathcal{L}\rho
\end{equation}
where $\mathcal{L} = \partial_{q_i}H\,\partial_{p_i}-\partial_{p_i}H\,\partial_{q_i}$ is the generator of the Perron-Frobenius presented in the Sec.\ \ref{Sec: KvN}.
Using the notation of Sec.\ \ref{Sec: KvN}, the generator takes the form $\mathcal{L}= -\mathbf{F}.\mathbf{\nabla}$.

\subsection{Ehrenfest equation}

In this statistical description, the value of an observable $f$ is defined through its average value \textit{i.e.}

\begin{equation}
    \langle f\rangle_\rho (t) = \int_\Omega f(\textbf{q}, \textbf{p})\rho(\textbf{q}, \textbf{p},t) d\textbf{q} d\textbf{p}.
    \label{Eq: Statistical observable}
\end{equation}

Assuming that the observable is static in time, the mean value of $f$ evolves according to Eq.\ \refeq{Eq: Observable eq} -- we can prove the latter by deriving the mean-value and integrating by part after injecting Eq.\ \refeq{Eq: Liouville}.
The latter is often referred to as Ehrensfest theorem
\begin{equation}
    d_t \langle f\rangle_\rho = \langle \{f, H\} \rangle_\rho.
    \label{Eq: Ehrenfest}
\end{equation}
In order to recover exactly Eq.\ \ref{Eq: Observable eq}, the density function has to be a Dirac delta function \textit{i.e.} the system should be well resolved in phase space.

Note that, an extra assumption has to be made for the existence of canonical variables which are analogous discussions as the one presented in Sec.\ \ref{Sec: KvN}.

\section{Proof of different bounds}
\label{App: Ehr bound}
\subsection{Bound}
This subsection aims at proving Eq.\ \ref{Eq: Ehrenfest bound}. 
We  start from a function $x_i: \Omega\rightarrow\mathbb{R}$ defined as $f_i(\textbf{x})=x_i$ whose average value over the density $\rho$ should match the solution $\tilde{x}_i(t)$.
Taking the difference between the original ODE equation (Eq.\ \ref{Eq: ODE classic}) and its Ehrenfest counterpart (Eq.\ \ref{Eq: Ehr Eq}), we get
\begin{equation}
    \frac{d(\langle x_i\rangle_\rho-\tilde{x}_i)}{dt}=\langle F_i(\textbf{x})\rangle_\rho- F_i(\tilde{\textbf{x}}) =\langle F_i(\textbf{x})- F_i(\tilde{\textbf{x}})\rangle_\rho
\end{equation}

Since $F_i$ is assumed to be locally $K_i$ Lipschitz, we can write\footnote{To be exact, the open set over which the locally Lipschitz is defined must contain the part where $\rho(\textbf{x},t)$ is not null. We could also considered $K_i$ to be a global Lipschitz constant.} (assuming the norm over $\Omega$ is the usual 2-norm).
\begin{equation}
    \langle |F_i(\textbf{x})- F_i(\tilde{\textbf{x}})|\rangle_\rho\leq K_i \langle |\textbf{x}-\tilde{\textbf{x}}|_2\rangle_{\rho}
\end{equation}
The prediction of the bound can be made in a slightly more tractable way by using the Cauchy Schwartz inequality on the right hand side:
\begin{equation}
    \left|\frac{d(\langle x_i\rangle_\rho-\tilde{x}_i)}{dt}\right|\leq K_i \sqrt{\sum\limits_{i=1}^d\text{Var}_{\rho(t)}(x_i)+|\langle \textbf{x}\rangle_\rho-\tilde{\textbf{x}}(t)|^2_2}
\end{equation}
\subsection{Covariance bound}
\label{Subapp: Reynolds bound}
Using Ehrenfest on the function $f_{i,j}:\Omega\rightarrow(x_i-\tilde{x}_i)(x_j-\tilde{x}_j)$ where $\tilde{\Sigma}_{ij}=\langle f_{ij}\rangle_\rho$, the Ehrenfest equation for the empirical covariance matrix satisfies
\begin{align*}
\frac{d\tilde{\Sigma}_{ij}}{dt}
\simeq
\sum_{k}
\left(
J_{ik}\,\tilde{\Sigma}_{kj}
+
\tilde{\Sigma}_{ik}\,J_{kj}
\right)\\+
    O\!\left(
        \|D^2\mathbf{F}\|,
        \left\langle
        \|\mathbf{x}-\tilde{\mathbf{x}}\|^3
        \right\rangle_{\rho}
    \right).
\end{align*}
where $\textbf{J}=\nabla\textbf{F}=(\partial_{x_j}F_i)_{i,j}$. Hence, using triangular inequality and the fact that the derivative of the covariance Frobenius norm is upper bounded by the Frobenius norm of the covariance matrix derivative 
\[
\frac{d}{dt}\|\tilde{\boldsymbol{\Sigma}}\|_F
\leq
\left\|
\frac{d\tilde{\boldsymbol{\Sigma}}}{dt}
\right\|_F\leq 2\|\mathbf{J}\tilde{\mathbf{\Sigma}}\|_F\]

Noting that
\[\|\mathbf{J}\tilde{\mathbf{\Sigma}}\|_F^2=\sum_{i,j,k}\left|\partial_{x_i}F_k\tilde{\Sigma}_{kj}\right|^2\leq L^2\|\tilde{\Sigma}\|_F^2 \]
where L is defined as $L:=\max_i \|\nabla F_i\|$.
By integrating, we get
\[
\|\tilde{\boldsymbol{\Sigma}}(t)\|_F
\leq
\|\tilde{\boldsymbol{\Sigma}}(t_0)\|_F
e^{2L |t-t_0|}.
\]

\section{Empirical approach on the Non Phase mixing criterion}
\label{App: Non-Phase Mixing}

In this section, the numerical criterion \ref{Eq: Numerical criterion} is derived more generally from basic consideration of the numerical scheme and can be generalized over the simulation.
As explained in Sec.\ \ref{Sec: ODE Lim}, this approach aims at bounding how much the gaussian is deformed along one Trotter step along the $x_i$ direction.
Fig.\ \ref{Fig: Gaussian shear} shows such an evolution under two typical flow behavior.
To simplify the discussion, the main result in this section are written for the multivariate Gaussian case with covariance matrix $\sigma I_d$ (where $\sigma$ is a scalar and $I_d$ the identity matrix in dimension $d$).

To quantify the picture, let us introduce the quantity:
\begin{equation}
    \Delta_\sigma x_i:= \max_{\boldsymbol\xi\in \mathcal{B}^d} |x_i(\textbf{x}+\sigma\boldsymbol{\xi},\Delta_t)-x_i(\textbf{x},\Delta_t)|\,,
\end{equation}
where $x_i(\textbf{x},t)$ is the value of $x_i$ after a translation along the $x_i$ axis of time $t$ starting from the initial position $\textbf{x}$ and $\mathcal{B}^d$ is the $d$-dimensional unit sphere for the 2-norm.
Empirically, $\Delta_\sigma x_i$ is a measure the maximum shear of a multivariate Gaussian after a Trotter step along axis $i$.
Geometrically, $\Delta_\sigma x_i$ measure the maximal red arrow on Fig.\ \ref{Fig: Gaussian shear}.b.

% This criterion stability in terms of mean standard deviation at initial time is motivated by the knowledge of Eq.\ \ref{Eq: Ehrenfest bound} and the fact that the Liouville theorem conserves the volume over time.
% %
% For the rest of this section, the focus is put on the multivariate gaussian case with uniform standard deviation over all axis $\sigma$

From a Taylor, expansion on the evolution of $x_i$ assuming $\sigma$ small compared to the typical variation scale of $F_i$, one can get in arbitrary dimension(with use of Einstein sum convention over $j$ and $k$ indices) :
\begin{align*}
    \Delta_\sigma x_i &\simeq\Delta t\max_{\xi \in \mathcal{B}^d}(F_i(\textbf{x}+\sigma\boldsymbol\xi)-F_i(\textbf{x}))\\
    &\simeq\Delta t \max_{\boldsymbol\xi \in \mathcal{B}^d}(\sigma\xi_j\partial_{x_j}F_i|_{\textbf{x}}+\frac{\sigma^2\xi_j\xi_k}{2}\partial_{x_j}\partial_{x_k}F_i|_{\textbf{x}})\\
    &\simeq\Delta t\,\sigma|\!|\nabla F_i|\!|_+\frac{\sigma^2\Delta t}{2}\|D^2F_i|_{\textbf{x}}\|_2
\end{align*}
where the canonical matrix 2 norm is defined in implicit summation
\[
\|D^2F_i|_{\mathbf{x}}\|_2
:=
\max_{\boldsymbol{\xi}\in\mathcal{B}^d}
\left|
\xi_j\xi_k
\partial_{x_j}\partial_{x_k}F_i|_{\mathbf{x}}
\right|\leq\|D^2F_i|_{\mathbf{x}}\|_F,
\]

To relate this local estimate to Criterion~\ref{Eq: Pract crit}, we normalize the maximal deformation by the characteristic width of the density. For an isotropic Gaussian with covariance matrix proportional to the identity, this length scale is $\sigma$; for a general covariance matrix, it may be replaced by the effective width $\sigma_F$. Since the flow is volume-preserving, stretching along some directions is compensated by compression along others. Requiring the relative deformation over one time step to remain small ensures that the flow is approximately affine across the support of the density.

Using the estimate derived above, we obtain
\begin{equation}
    \frac{\Delta_\sigma x_i}{\sigma}
    \lesssim
    \Delta t\,\|\nabla F_i(\mathbf{x})\|
    +
    \frac{\sigma\Delta t}{2}
    \|D^2F_i(\mathbf{x})\|_F .
\end{equation}
The two terms describe distinct contributions to the deformation. The first corresponds to the linear stretching generated by the Jacobian and imposes a condition on the time step,
\begin{equation}
    \Delta t\,\|\nabla F_i(\mathbf{x})\|\ll 1.
\end{equation}
The second quantifies the nonlinear deformation induced by the curvature of the vector field. Once the time step has been chosen to control the linear contribution, comparing the nonlinear and linear terms yields the Ehrenfest--Reynolds number $R$ introduced in Sec.~\ref{Sec: ODE Lim} and thus bounding the second term is equivalent to the condition $R\ll1$. Thus, requiring both contributions to remain small provides a sufficient local condition for Criterion~\ref{Eq: Pract crit}.

\section{Asymptotic estimate of the number of measurement}
\label{App.: Bayes}
\subsection{Sampling estimates}
% The reasoning behind using Bayesian theory is the following: Quantum computation will give you outcome, how do you interpret the result since there is a probabilistic outcome?
% %
% Indeed, the outcome of the method is a direct sampling of the distribution $\rho$ according to Born rule.
% %
% As a result, the probabilistic outcome makes the result hard to interpret and the benchmarking of the code is thus difficult.
% %
% This issue is usually answered in Monte Carlo community where sampling distribution is important.
% %
To benchmark the ODE solver, one needs to know the number of measurement required to get an estimate of an average value as defined in Sec.\ \ref{Sec:Measurement}.
In this section, an asymptotic estimate of the number measurement is proposed using direct sampling or Bayes theory based on Ref.\ \cite{Cocco2022}.
The ODE solver is assumed to preserve the Gaussian shape of the density function.
Since each part of the register is probed, the problem will be simplified to sampling one quantity $x$ following a Gaussian distribution 
\begin{equation}
    \mathbb{P}(x)= \frac{1}{\sqrt{2\pi \sigma^2 }} e^{\frac{-(x-\langle x\rangle)^2}{2\sigma^2}}
\end{equation}
The latter correspond to a distribution obtained after using Born rules on the resulting density assuming it is gaussian. 
It is further assumed that the size $\sigma$ of the Gaussian is known since quantum measurement can be designed to measure it. 

The previous problem can be rewritten as following, after $M$ iteration of the quantum computer, a list of $M$ samples of the position will be obtained.
The aim is to find the conditions necessary to get an estimate of the mean of the Gaussian $\langle x\rangle$.
One would first assume that the likelihood for the data to be estimated is 
\begin{equation}
    \mathbb{P}(x_m|\tilde{x}) = \frac{1}{\sqrt{2\pi \sigma^2}} e^{\frac{-(x-\tilde{x})^2}{2\sigma^2}}.
\end{equation}

Assuming measurement independent, one can write the Bayes rule for the probability of estimating $\tilde{x}$ given our set of measurements $(x_m)_{m\in[\![1,M]\!]}$

$$\mathbb{P}(\tilde{x}| (x_m)_{m\in[\![1,M]\!]}) = \frac{\mathbb{P}(\tilde{x})\, \prod_m\mathbb{P}(x_m|\tilde{x})}{\int d\tilde{x} \prod_m\mathbb{P}(x_m|\tilde{x})}.$$

Due to the a priori lack of information on $\tilde{x}$ the prior $\mathbb{P}(\tilde{x})$ is taken uniform.
Other prior $\mathbb{P}(\tilde{x})$ can be defined but would require further analysis or assumptions \cite{Cocco2022}.

The likelihood $ \prod_m\mathbb{P}(x_m|\tilde{x})$ can be written as 
\begin{eqnarray*}
    \prod_m\mathbb{P}(x_m|\tilde{x}) = \exp\left( M \frac{1}{M}\sum\limits_{m=1}^{M} \ln \mathbb{P}(x_m|\tilde{x}) \right).
\end{eqnarray*}

Since the $x_m$ are sampled over the distribution $\mathbb{P}(x_m|\langle x\rangle)$ with the correct estimate, the asymptotical likelihood takes the form
\begin{eqnarray*}
    \prod_m\mathbb{P}(x_m|\tilde{x}) = \exp(-M S_c(\tilde{x}, \langle x\rangle))\\
    S_c(\tilde{x},\langle x\rangle) =-\int_0^1 dx\,  \mathbb{P}(x|\langle x\rangle)\ln \mathbb{P}(x|\tilde{x})
\end{eqnarray*}

where $S_c$ is the cross-entropy.
If $\tilde{x} = \langle x\rangle$, the cross entropy becomes the standard entropy of the probability distribution.

To simplify the calculation and since the Gaussian size is small compared to the domain size ($\sigma<\!< 1$), the integral bound is extended to the real axis. 

For two Gaussian distribution with the same deviation $\sigma$, the cross entropy is
\begin{eqnarray*}
   S_c(\tilde{x}, \langle x\rangle) &= -\int_{-\infty}^{+\infty} dx\,  \mathbb{P}(x|\langle x\rangle)\ln \mathbb{P}(x|\tilde{x})\\
   & = \frac{1}{2}+\frac{(\tilde{x}-\langle x\rangle)^2}{2 \sigma^2}.
\end{eqnarray*}

As a result, the probability for the estimate $\tilde{x}$ to be more than $\epsilon$ away from the real value $\langle x \rangle$ is scaling as

\begin{eqnarray*}
    \mathbb{P}(|\tilde{x}-\langle x \rangle|\geq \epsilon) \propto \int_{|\tilde{x}-\langle x \rangle|\geq \epsilon}d\tilde{x}\, e^{-M\frac{(\tilde{x}-\langle x\rangle)^2}{2 \sigma^2}}\sim e^{-\frac{\epsilon^2}{2\sigma^2}}.
\end{eqnarray*}

The latter scaling is due to the large number limit which narrows the Gaussian width of this probability.
The number of measurement necessary to have $99\%$ chance to estimate the average value within an error $\epsilon$ is scaling as
\begin{equation*}
    M \sim \frac{\sigma^2}{\epsilon^2}\, \ln \frac{1}{0.01} = O\left(\frac{1}{\epsilon^2}\right)
\end{equation*}

As a result, we can estimate the number of measurement needed to estimate the characteristics of a characteristic observable.
\subsection{Bernouilli variables} 

Similarly for a Bernouilli statistics which true distibution is given by $\mathbb{P}_0$, the cross entropy is 
\begin{equation}
S_c(\tilde{\mathbb{P}}_0,\mathbb{P}_0)=-\mathbb{P}_0\ln\tilde{\mathbb{P}}_0-(1-\mathbb{P}_0)\ln(1-\tilde{\mathbb{P}}_0)
\end{equation}
upon Taylor expanding around $\tilde{\mathbb{P}}_0=\mathbb{P}_0+\epsilon$
with $\epsilon \ll p$:
\begin{equation}
S_c(\mathbb{P}_0+\epsilon,\mathbb{P}_0)=S_c(\mathbb{P}_0,\mathbb{P}_0)+\frac{\epsilon^2}{2\,\mathbb{P}_0(1-\mathbb{P}_0)}+O(\epsilon^3)
\end{equation}
Using as before the marginalization over the posterior tail at leading order in $\epsilon$:
\[
\mathbb{P}\!\left(|\tilde{x}-\langle x\rangle|\geq \epsilon\right)\sim e^{-\frac{M\,\epsilon^2}{2\mathbb{P}_0(1-\mathbb{P}_0)}}.
\]
which in turns gives the number of measurement necessary to have $99\%$ chance to estimate the average value within an error $\epsilon$ is scaling as
\begin{equation}
    M\simeq \frac{2\mathbb{P}_0(1-\mathbb{P}_0)}{\epsilon^2}\ln(0.01).
\end{equation}

The latter also highlights that a deviation from $\mathbb{P}_0=0$ or $\mathbb{P}_0=1$ can be inferred in few measurements.

\subsection{Probabilistic Grover inference}

Grover-type algorithms allow we to reduce the asymptotic measurement complexity from
$O(1/\epsilon^2)$ to $O(1/\epsilon)$ for the estimation of Bernoulli probabilities
\cite{Brassard2002}. In the present setting, the goal is to use this mechanism to infer an effective angle associated with the averaged observable encoded in the phase function $\theta(Q,P)$.
\begin{figure}
    \centering
    \begin{quantikz}
\lstick[wires=2]{$\ket{\rho}(t)$} & \qwbundle{n}  &\gate[wires=2]{e^{i\theta(X_1,X_2)}} & \qw \\
&\qwbundle{n} & \qw & \qw  \\
\lstick{$\ket{0}$} & \gate[wires =1]{H} & \ctrl{-1} & \gate[wires =1]{H} & \gate[wires =1]{e^{i\sigma_x \theta_m}} & \meter{\hat{Z}}
\end{quantikz}
    \caption{
Measurement circuit used for the probabilistic Grover inference in a two-dimensional phase space. The registers $\ket{X_1}$ and $\ket{X_2}$ encode the phase-space grid, $U_t$ denotes the KvN time-evolution operator, and the diagonal phase unitary encodes the observable-dependent angle $\theta(X_1,X_2)$. The final rotation by $\theta_m$ shifts the measurement reference before measuring the ancilla in the $\hat Z$ basis.
}
    \label{fig: Grover}
\end{figure}
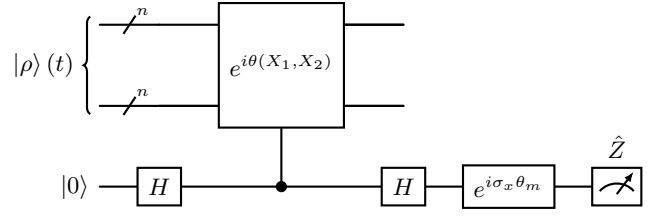
As a result, the probability of measuring 0 on the measurement qubit can be written as

\begin{align}
\mathbb P_0
&= \frac{1}{\mathcal{N}}
\sum_{(i,j)\in[\![1,2^n]\!]^2}
\cos^2\!\left(\theta(Q_i,P_j)-\theta_m\right)
\rho(Q_i,P_j,t)
\\
&=
\frac{1}{2}
+
\frac{1}{2}
\left\langle
\cos\!\left(2(\theta-\theta_m)\right)
\right\rangle_\rho
+
O(2^{-n}).
\end{align}
where the last equality follows from the same Riemann approximation argument used in Sec.~\ref{Sec:Measurement} and $\mathcal{N}$ is defined as
$$\mathcal{N}=\sum_{(i,j)\in[\![1,2^n]\!]^2}\rho(Q_i,P_j,t).$$

$\mathbb P_0$ can be related to an effective Grover angle by expanding around the averaged angle $\langle\theta\rangle_\rho$.

\begin{align}
\mathbb{P}_0
&=\frac{1}{2}+\frac{1}{2}
\cos\!\left(
2(\langle\theta\rangle_\rho-\theta_m)
\right)
\left(
1-2\,\operatorname{Var}_\rho(\theta)
\right)
+
O\!\left(
\langle
|\theta-\langle\theta\rangle_\rho|^3
\rangle_\rho
\right),\label{ApEq: Probability of quantum counting}\\
&\simeq \cos^2(\langle\theta\rangle_\rho-\theta_m)-\cos\!\left(
2(\langle\theta\rangle_\rho-\theta_m)
\right)\,\operatorname{Var}_\rho(\theta)
\end{align}
where

\[
\operatorname{Var}_\rho(\theta)
=
\left\langle
(\theta-\langle\theta\rangle_\rho)^2
\right\rangle_\rho.
\]

Assuming that $\theta$ is twice continuously differentiable and that $\rho$ remains sufficiently localized, the variance admits the approximation

\begin{equation}
\operatorname{Var}_\rho(\theta)
=
\nabla\theta(\langle\mathbf{x}\rangle_\rho)^\top
\mathbf{\Sigma}
\nabla\theta(\langle\mathbf{x}\rangle_\rho)
+
O(\|\mathbf{\Sigma}\|^2).
\end{equation}
which is independent from the inferring process.

To extract $\langle\theta\rangle_\rho$ we introduce $\theta_m=m\theta_0$ where $m$ is a number of iteration of this amplitude estimation algorithm and $\theta_0=\epsilon$ and assuming $\operatorname{Var}_\rho(\theta)$ fixed by the quantum state.
With this formalism, we can apply directly an amplitude amplification in Ref.\ \cite{Brassard2002} or more exactly the proof developed in the Theorem 3.1 of Ref.\ \cite{Aaronson2020}.
However the precision $\epsilon$ at which we estimate $\langle\theta\rangle_\rho$ is at best $\operatorname{Var}_\rho(\theta)=O(\|\mathbf{\Sigma}\|\,\|D^2\theta\|)$ .
We can see the latter in Eq.\ \ref{ApEq: Probability of quantum counting} by introducing $\cos2\theta_{err}=(1-2\operatorname{Var}_\rho(\theta))$ upon the small angle assumption $(\langle\theta\rangle_\rho-\theta_m)\ll1$ which yields to $\mathbb{P}_0\sim\cos^2(\langle f\rangle_\rho-m\epsilon+\theta_{err})$.

As a result, using $\theta = f$ amplitude estimation algorithm estimates $\langle f\rangle_\rho$, the theorem 3.1 of Ref.\ \cite{Aaronson2020} we can estimate of $\langle f\rangle_\rho+\theta_{err}$ at a precision $\epsilon$ with a probability of failure $\delta$ with \[M=O\left(\frac{1}{ \cos^2(\langle f\rangle+\theta_{err})\rangle_\rho \,\epsilon}\ln\delta^{-1}\right)\]

More refined observable-estimation strategies based on block encodings and amplitude estimation, such as those developed in Ref.\ ~\cite{Rall2020}, could improve this measurement step.

\section{Lokta-Volterra system basics}
\label{Ap: LV}
The 2 species Lokta-Volterra system is a dynamical system with two species population: the prey represented by $x_1(t)$ and the predator represented by $x_2(t)$. 
The prey population unconditionally grows with the rate $\alpha$ and in presence of predator decays with a rate $\beta x_2(t)$.
Similarly, the predator population unconditionally decays with a rate $\delta$ and increase with a rate $ \gamma x_1(t)$.
The Lokta-Volterra system is important in plasma physics as the system appears in zonal flow turbulence and the Kim-Diamond model.

The differential equation system for Lokta-Volterra thus reads
\begin{subequations}
\begin{equation}
    \partial_t x_1(t) = \alpha x_1(t) -\beta x_1(t) x_2(t)
\end{equation}
    \begin{equation}
        \partial_t x_2(t) = -\delta x_2(t) +\gamma x_1(t) x_2(t)
\end{equation}
\label{Eq: Lokta}
\end{subequations}

The following system presents two stable points: an unstable trivial one at $x_1 = 0, \,x_2=0$ and another stable at $x_1 = \delta/\gamma, \,x_2=\alpha/\beta$ that is an attractor.

Since $(0,0)$ is an unstable attractor, one can rephrase  the system in the more convenient form 

\begin{subequations}
\begin{equation}
    \partial_t \ln x_1(t) = \alpha -\beta \exp(\ln x_2(t))
\end{equation}
    \begin{equation}
        \partial_t \ln x_2(t) = -\delta  +\gamma \exp(\ln x_1(t))
\end{equation}
\label{Eq: Lokta cano}
\end{subequations}

We notice that the latter system is Hamiltonian in the $\ln x_1, \,\ln x_2$ variables with the conserved quantity

\begin{equation}
    H(\ln x_1,\ln x_2) = \delta \ln x_1  -\gamma  \exp\ln x_1 +\alpha \ln x_2 - \beta \exp(\ln x_2)
    \label{Eq: LV Ham}
\end{equation}

For convenience, the system is expressed in its canonical variables $q = - \ln x_1$ and $p = -\ln x_2$ where the minus signs have no physical relevance. 
In these variables, the system becomes explicitly Hamiltonian:

\begin{subequations}
\begin{equation}
    H(p, q) = -\delta q  -\gamma  \exp(-q ) -\alpha p - \beta \exp(-p)
\end{equation}
\begin{equation}
    d_t q = -\alpha +\beta \exp(-p) = \partial_p H
\end{equation}
    \begin{equation}
        d_t p = +\delta  -\gamma \exp(-q) =-\partial_q H.
\end{equation}
\label{Eq: LV Ham can}
\end{subequations}

In these variables, we define a Canonical phase space structure with the Poisson Bracket geometry $\{A,B\} = \partial_q A \partial_p B - \partial_q B\partial_pA$.
In this canonical coordinate, the Perron Frobenius generator reads
\begin{equation}
    \partial_t \rho = (-\delta+\gamma e^{-q})\partial_p\rho - (\alpha -\beta e^{-p})\partial_q\rho
\end{equation}

And the Liouvillian of the automata for a stencil 2 is 
\begin{equation}
    \mathcal{\hat{L}}_{LV}= (-\delta+\gamma e^{-\hat{Q}\Delta q})\,\frac{\hat{\mathcal{S}}_p-\mathcal{S}^\dagger_p}{2\Delta p} + (\alpha -\beta e^{-\hat{P}\Delta p })\,\frac{\mathcal{S}_q-\hat{\mathcal{S}}^\dagger_q}{2\Delta q} 
\end{equation}

\section{Walsh operator}
\label{App: Walsh}

In this section, we will show how to encode a diagonal unitary based on Ref.\ \cite{WGM14}.
For simplicity we consider $f$ a function of one variable defined on $[0,1]$.
The aim is to have an efficient representation of the discretized function $f$ defined on a regular grid.
A function $f$ is valuated on $2^{N}$ points and we note $f_n := f(n/2^N)$ for $n \in [\![0, 2^N-1]\!]$.
The aim is to represent the $f_n$ as a Walsh serie $f_n = f(n/2^N) = \sum_j a_j w_j(n/2^N)$

The Walsh functions $w_j$ are defined for an integer $j$ decomposed on the binary basis $j=\sum_i j_i 2^i$  and a coordinate $x\in [0,1]$ which can be represented by its dyatic expansion $x = \sum_i x_i \frac{1}{2^i}$.
The Walsh function reads 
\begin{eqnarray*}
    w_j : [0,1]&\mapsto\{1,-1\}\\
       x&\mapsto (-1)^{j.x}
\end{eqnarray*}
where $j.x = \sum_i j_i x_i$ is the binary scalar product.
The advantage of Walsh transform is purely computational as the Walsh transform can be seen as a binary Fourier transform and is sometimes referred as the Walsh-Fourrier transform in the continuous case.
As the Fourier transform, a discrete Walsh transform can be developed.
The latter is often refer to as Walsh-Hadamard transform.
In a tensor form, the Walsh-Hadamard transform is represented by a tensor product of single Hadamard matrix convenient for circuit depth in comparaison to QFT.

Expanded on $2^N$ points, the Walsh series of the discretized function $f_n$ can be represented exactly with $2^N$ Walsh coefficent.
In this case, the Walsh transform reads
\begin{eqnarray*}
    &a_j = \frac{1}{2^N}\sum_{n=0}^{2^N-1} f_n w_j(n/2^N)\\
    &f_n = \sum_{j=0}^{2^N-1}a_j w_j(n/2^N)= \sum_{j=0}^{2^N-1} a_j \prod_{i=0}^{N-1} (-1)^{x_i j_i}.
\end{eqnarray*}

The discretized $f$ can be represented as an operator that can be written as:
\begin{equation}
    \hat{f} = \sum_{j=0}^{2^N-1} a_j \bigotimes_{i=0}^{N-1} \hat{\sigma}_{z}^{j_i}
\end{equation}

On the discrete grid state $\ket{n}$, the action of the diagonal unitary $e^{i\hat{f}}$ is $e^{i\hat{f}}\ket{n} =e^{if_n}\ket{n} $.

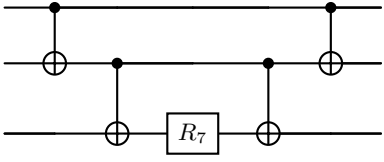
\begin{figure}
    \centering
    \begin{quantikz}
        \qw      & \ctrl{1} & \qw      & \qw & \qw      & \ctrl{1} & \qw \\
        \qw      & \targ{}  & \ctrl{1} & \qw  & \ctrl{1} & \targ{}  & \qw \\
        \qw      & \qw      & \targ{}  & \gate{R_7} & \targ{}  & \qw      & \qw
    \end{quantikz}
    \caption{Quantum circuit implementing a controlled $e^{ia_7 \bigotimes_{i=0}^{2} \sigma_{z}}$ gate with $R_z (a_7) =R_7$. This figure is extracted from Ref.\ \cite{WGM14}.}
    \label{fig:controlled_R7}
\end{figure}

\begin{figure*}
    \centering
    \begin{quantikz}
        \lstick{$\ket{n_1}$} & \gate{R_1} & \ctrl{1} & \qw & \ctrl{1} & \qw & \qw & \qw & \ctrl{2}&\qw & \qw & \qw & \ctrl{2}&\qw&\qw\\
        \lstick{$\ket{n_2}$} & \qw        & \targ{}  & \gate{R_3} & \targ{}  & \gate{R_2}  & \ctrl{1} & \qw &\qw  &\qw&\ctrl{1}& \qw& \qw &\qw  &\qw\\
        \lstick{$\ket{n_3}$} & \qw       &\qw & \qw      & \qw      & \qw    & \targ{} & \gate{R_6} & \targ{}  & \gate{R_7} & \targ{} & \gate{R_5}&\targ{}&\gate{R_4}&\qw
    \end{quantikz}
    \caption{Optimized quantum circuit implementing the Paley-ordered Walsh operators on 3 qubits. This figure is extracted from Ref.\ \cite{WGM14}.}
    \label{fig:optimized_walsh}
\end{figure*}
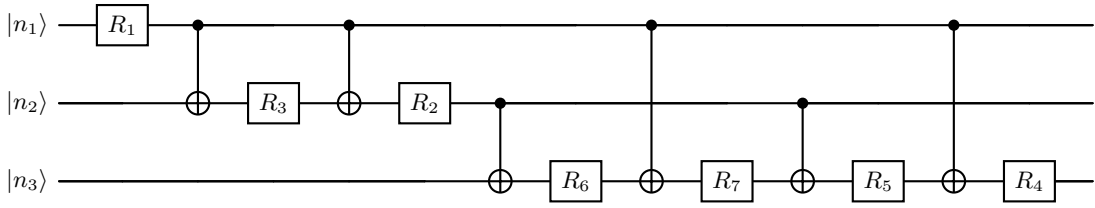

The diagonal unitary operator is expressed as a product of gate of the type $e^{ia_j \bigotimes_{i=0}^{N-1} \sigma_{z}^{j_i}}$ which is a generalized $R_z$ gate.
This gate can be written as a single gate $R_z (a_j) =R_j$ and a CNOT layer as presented in Fig.\ \ref{fig:controlled_R7}.
A full Walsh decomposition on 3 qubits are represented  Fig.\ \ref{fig:optimized_walsh}.

This decomposition could be further simplify by keeping only the relevant Walsh coefficient $a_j$ as suggested by Ref.\ \cite{Zylberman2024}.
Scalings and details about the approach can be found in Ref.\ \cite{WGM14}.

\section{Initial condition treatment}
\label{App: State Prep}

One can prepare a state efficiently using the circuit presented in Ref.\ \cite{Zylberman2024}.
The main idea is to prepare the state
\begin{eqnarray*}
    \ket{\Psi} = \frac{1}{\sqrt{2}}\sum_{X,P} &(e^{i f(X,P)} + e^{-i  f(X,P)} )\ket{X,P,0}\\&+(e^{i  f(X,P)} -e^{-if(X,P)} )\ket{X,P,1}
\end{eqnarray*}
If the ancilla registered is measured as $1$ (the measurement is short hand as $\langle 1|\Psi\rangle$), the state after measurement is 

\begin{eqnarray*}
	\ket{\psi}&= \frac{\langle 1|\Psi\rangle}{\sqrt{|\langle 1|\Psi\rangle|}}\\
	&=\frac{\sum_{X,P} (e^{i  f(X,P)} -e^{-i f(X,P)} )}{\sqrt{\sum_{X,P} \sin^2( f(X,P))}}\\ 
	&\sim \sum_{X,P} \frac{f(X,P)}{||f(X,P)||_2} \ket{X,P}
\end{eqnarray*}

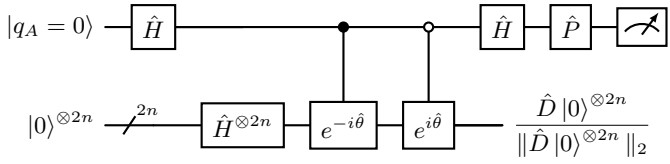
\begin{figure}
    \centering
    \begin{quantikz}[row sep=0.7cm, column sep=0.35cm]
        \lstick{$\ket{q_A=0}$}
        & \gate{\hat H}
        &\qw
        & \ctrl{1}
        & \octrl{1}
        & \gate{\hat H}
        & \gate{\hat P}
        & \meter{} \\
        \lstick{$\ket{0}^{\otimes 2n}$}
        & \qwbundle{2n}&\gate{\hat H^{\otimes 2n}}
        & \gate{e^{-i\hat\theta}}
        & \gate{e^{i\hat\theta}}
        & \rstick{$\displaystyle\frac{\hat D\ket{0}^{\otimes2n}}{\|\hat D\ket{0}^{\otimes 2n}\|_2}$}\qw
    \end{quantikz}
    \caption{Quantum circuit for block-encoding the non-unitary diagonal operator $\hat F=\operatorname{diag}(f_0,\ldots,f_{2^{2n}-1})$. The phase operator is defined by $\hat\theta=\arcsin(\hat F/f_{\max})$, where $f_{\max}=\max_j|f_j|=1$. Upon successful postselection of the ancilla, the system register is prepared in the normalized state $\hat D\ket{\psi}/\|\hat D\ket{\psi}\|_2$. Adapted from Ref.~\cite{Zylberman2024}.}
    \label{fig:block_encoding_state_preparation}
\end{figure}

Probability details and scaling are proposed in Ref.\ \cite{Zylberman2024}.
% \section{Discussion on the Unitarity of Koopman Operator}

% Constructing an invariant measure associated with a given flow is, in general, a non-trivial problem. While the existence of such a measure can be established under suitable assumptions, its explicit form depends on global properties of the dynamical system and cannot be readily inferred from the local ODE formulation. 
% %
% Even in the presence of conserved quantities, identifying a representation in which the invariant measure takes a simple form remains challenging and often requires non-trivial transformation as highlighted in the Lotka-Volterra system in App.\ \ref{Ap: LV}. In practice, determining these transformations is problem-dependent and may not be feasible in general.
% %

\end{document}